# Microscale optoelectronic reservoir networks of halide perovskite for in-sensor computing

Jeroen J. de Boer[1], Agustin O. Alvarez[1], Moritz C. Schmidt[1], Bruno Ehrler[1,*]
[1]LMPV-Sustainable Energy Materials Department, AMOLF, 1098 XG, Amsterdam, the Netherlands
*Correspondence: b.ehrler@amolf.nl

## Abstract

Physical reservoir computing is a promising framework for efficient neuromorphic in and near-sensor computing applications. Here, we demonstrate a multimodal optoelectronic reservoir network based on halide perovskite semiconductor devices, capable of processing both voltage and light inputs. The devices consist of micrometer-sized, asymmetric crossbars covered with a $MAPbI_3$ perovskite film. In a network, we simulate the performance by transforming MNIST images and videos based on the N-MNIST dataset using 4-bit inputs and training linear readout layers for classification. We demonstrate multimodal networks capable of processing both voltage and light inputs, reaching mean accuracies up to 95.3 ± 0.1% and 87.8 ± 0.1% for image and video classification, respectively. We observed only minor deterioration due to measurement noise. The networks significantly outperformed linear classifier references, by 3.1% for images and 14.6% for video. We show that longer retention times benefit classification accuracy for single-mode networks, and give guidelines for choosing optimal experimental parameters. Moreover, the microscale device architecture lends itself well to further downscaling in high-density sensor arrays, making the devices ideal for efficient in-sensor computing.

## Introduction

While upscaling of neural networks has resulted in impressive increases in their capabilities, it has also led to an exponential rise in energy consumption.[1] Novel neuromorphic hardware neural networks inspired by the brain are an appealing, more energy-efficient alternative.[2,3] Brain-inspired networks that process inputs close to the sensor are particularly interesting for reducing inefficient data transfer and power consumption.[4] Physical reservoir computing is a compelling approach for this purpose, as it leverages inherent device dynamics to preprocess inputs. In reservoir computing, a fixed dynamical system nonlinearly transforms inputs to increase linear separability,[5] after which a simple linear readout layer is used for classification.[6,7] As only the simple linear readout layer is trained, these networks are significantly easier to implement in hardware than neuromorphic networks with many complex trained layers.[2,8]

Physical reservoirs for in-sensor computing commonly apply the "single dynamical node" approach,[7] shown schematically in Figure 1a. A time-dependent input is fed into the device ("reservoir node"), which changes its state. The reservoir node has a volatile memory, so that its state depends on both the presented input and its history. The reservoir node states are collected over time, and a linear weighted sum is taken to yield the final output. For in-sensor computing, this approach is typically extended to arrays of reservoir nodes,[9] as illustrated schematically in Figure 1b. After the input **u**(t) is transformed, the final output **y** is obtained by taking a weighted sum of the reservoir states **x**. This can be implemented by running the readout reservoir states through a resistive or memristive device array (weights **W** in Figure 1b).[9] Optionally, the output can be collected for further processing on digital platforms. These networks do not require external memory if only the final device states are considered, strongly reducing inefficient data transport.[3]

A broad range of platforms is suitable for physical reservoir computing because the reservoir only requires a fixed nonlinear transformation of an input. Implementations include memristive devices,[9,10] photonic circuits,[11,12] and spintronic devices.[13] The ability of optoelectronic devices to process a broad range of inputs makes them particularly attractive for in-sensor reservoir computing.[14] These devices can process various multimodal optical and electronic inputs and combine them to increase classification accuracy.[15] Halide perovskites have excellent optoelectronic properties and are hence well-suited for reservoir computing. Next to strong light absorption, they feature complex and tunable transient behavior due to ion migration induced by a light or voltage bias.[16] Several reports have explored the use of halide perovskites for reservoir computing.[17–20] However, in these implementations, the reservoir was limited to detecting either a voltage or a light-based input, and multimodal sensing was not explored. Moreover, these reports did not address the difficulty of microfabrication of halide perovskite devices,[21] which is necessary for high-density integration.

Here, we address these gaps by implementing the microscale halide perovskite devices we recently developed as optoelectronic artificial synapses[22] for reservoir computing. After applying a voltage pulse, the devices output an ionic displacement current that decays on the seconds to hundreds of milliseconds timescale. Illuminating the device when a voltage is applied increases the current modulation. We investigate the linear separability of four-bit light and voltage inputs based on this volatile current. Based on the results, we construct in-sensor reservoir computing networks that transform and classify handwritten digits from images (MNIST) and video (modified N-MNIST), while accounting for experimentally measured noise. We obtain classification accuracies up to 95.3 ± 0.1% for image and 87.8 ± 0.1% for video datasets when combining both inputs in the same multimodal network, surpassing linear classifier references. When considering only one type of input, light-input networks outperformed networks based on voltage inputs. We show with simulations that this is due to the shorter retention time relative to the input frequency for voltage inputs. The simulations allow facile estimation of network accuracies based on the 4-bit input measurements, valuable when fine-tuning experimental parameters. They also show complementary transformations by the light and voltage networks, leading to the high accuracy of the multimodal networks. Our results demonstrate the potential of halide perovskite volatile devices for efficient, multimodal in-sensor computing.

# Results and Discussion

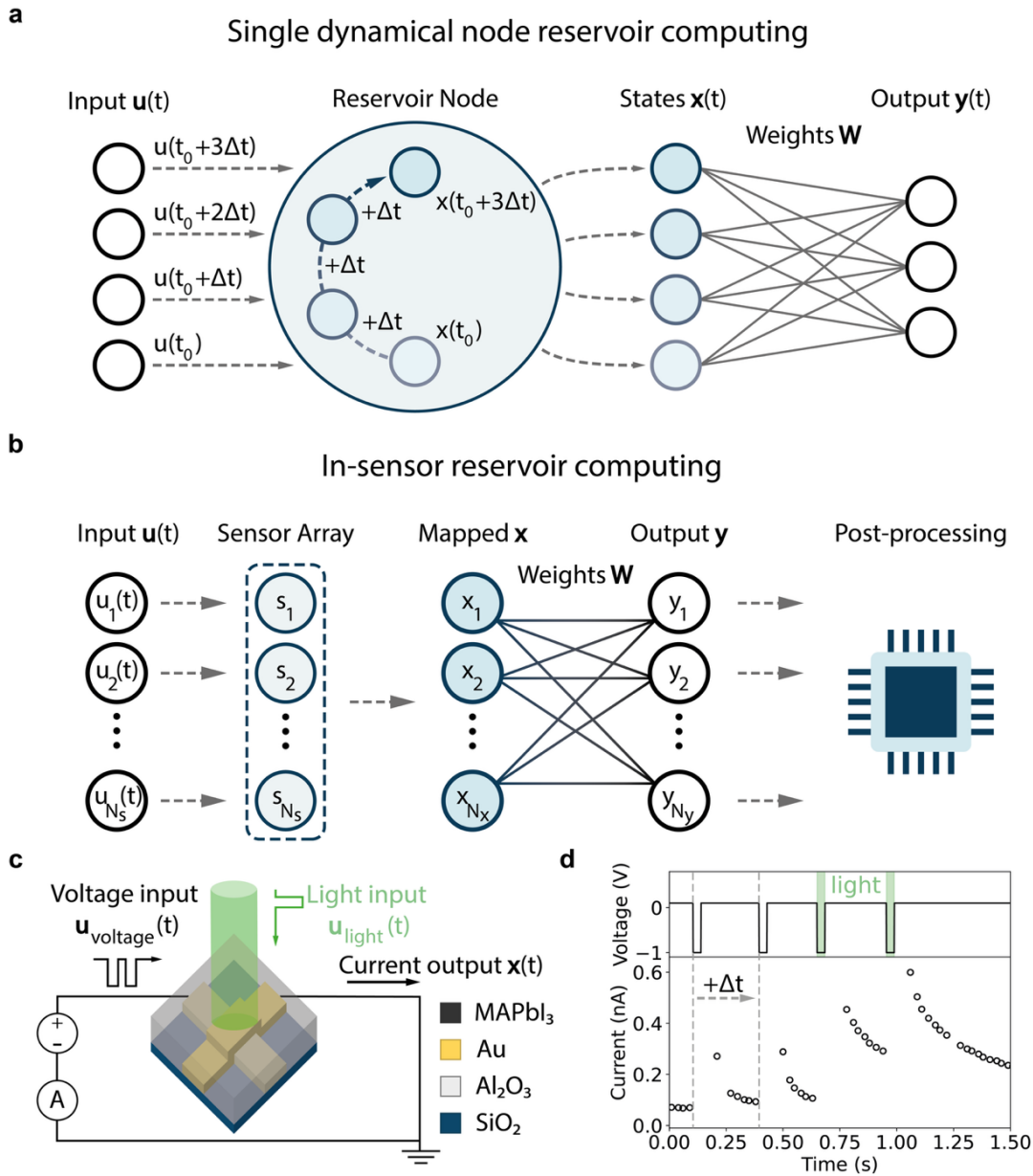

Figure 1. Reservoir computing with volatile artificial synapses. **(a)** Schematic representation of reservoir computing with a single dynamical node. This node can be implemented with a volatile device ("reservoir node") processing a time-encoded input, represented by input vector **u**(t). Each element of **u**(t) is an input at time t. The input changes some internal state of the node, resulting in a nonlinear transformation of the input. The output of the reservoir node at each point in time, represented by **x**(t), depends on the input at that time and the previous state of the node. The states collected in **x**(t) are read out by taking a weighted sum with the weight matrix **W**, giving the final output of the network, **y**(t). **(b)** Schematic of in-sensor reservoir computing with an array of volatile devices. Each device $s_n$ represents a reservoir node from **(a)** and receives a time-varying input $u_n(t)$, which is transformed into one or multiple states. The states of all nodes are collected in **x**. Similar to reservoir computing with a single dynamical node, the final output **y** is obtained by taking a weighted sum with the weight matrix **W**. Afterwards, the output can be collected for further processing. In both **(a)** and **(b)**, only the weights in **W** are adapted during training. **(c)** Schematic drawing of the volatile optoelectronic halide perovskite artificial synapse. The microscale device can process both voltage and light inputs (**u**(t)) and gives a current as an output (**x**(t)). **(d)** An example pulsed voltage measurement showing a slow current decay over hundreds of milliseconds after an applied pulse. A second pulse applied during the decay gives a higher current compared to the first pulse. Simultaneously illuminating the device when two more voltage pulses are applied increases the change in current. The currents measured during the pulses are omitted for clarity. The pulse and its following dwell time correspond to one timestep (Δt) in **(a)**.

A schematic image of the microscale halide perovskite device is shown in Figure 1c. The device consists of 2.5 µm-wide, back-contacted crosspoint electrodes of gold that sandwich an insulating $Al_2O_3$ layer. A $MAPbI_3$ layer is spin-coated over the crosspoint electrodes. We have previously developed this device architecture to prevent degradation of the perovskite layer during microfabrication.[22,23]

The example of a pulsed voltage and light measurement in Figure 1d shows how the device could be used as a reservoir node. Four -1 V pulses, corresponding to the four inputs at different times in Figure 1a, are applied to the device. The current, resembling the read-out states **x**(t), is measured continuously. Each combination of a -1 V pulse and the subsequent dwell time, during which the current is measured, represents one Δt timestep from Figure 1a. We use Δt to refer to the timesteps instead of the more conventional "τ" to avoid confusion with the characteristic decay time of the current.[24] Applying -1 V pulses results in a current that decays slowly over hundreds of milliseconds after each pulse. The current recorded after the second pulse (0.29 nA) is slightly larger than after the first pulse (0.27 nA). Higher current changes are obtained after the third (0.40 nA) and fourth (0.60 nA) -1 V pulses, during which the device was simultaneously illuminated. This difference demonstrates that the device's response to a new input, i.e. its read-out state, depends on both the input itself and the history of previous inputs, a requirement of reservoir computing. Crucially, both the voltage and light pulses affect the current in distinct ways.

Previously, we have shown that the current decay of this device after a voltage pulse is governed by a combined ionic drift and diffusion process.[22] Here, the decay follows the same ionic drift and diffusion processes, as follows from the fit in Figure S1. Drift-diffusion simulations in Figure S2 further corroborate that the transient current response is due to an ionic displacement current. The increase in current after each pulse is due to further accumulation of ions. Illumination during the -1 V pulse likely enhances the accumulation due to the higher ionic conductivity in light,[25] in line with our previous work.[22]

Using this volatile current for reservoir computing requires a linearly separable output for different inputs. We first investigate this linearity for voltage inputs without illumination. Next, we explore it for inputs combining voltage pulses with light inputs.

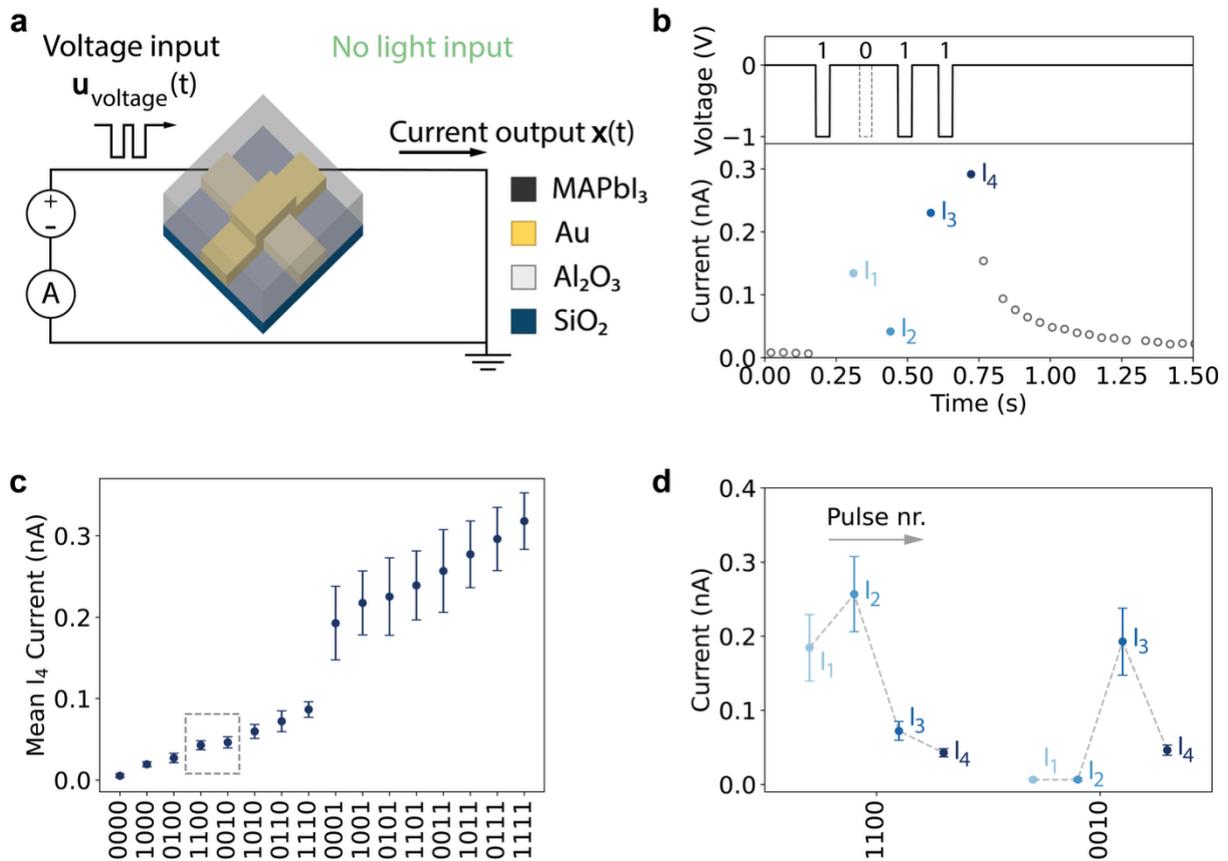

*Figure 2. Electronic measurements of four-bit voltage profiles. **(a)** Schematic of the measurements. A four-bit, -1 V pulsed voltage input is applied, and the resulting current is measured. **(b)** An example measurement of three -1 V pulses, shown in the top panel. The dotted line after the first pulse indicates a missing -1 V pulse for that Δt timestep. Hence, this voltage profile corresponds to a 1011 input. The bottom panel shows the measured currents. Currents after each of the four pulses used for further analysis ($I_1$, $I_2$, $I_3$, and $I_4$) are shown in blue. **(c)** Mean $I_4$ currents for each four-bit input. Output currents are higher for inputs with more -1 V pulses, and for inputs with pulses applied later in the 4-bit input. Some means are close to overlapping, such as those of the 1100 and 0010 inputs highlighted by the dashed gray box. **(d)** Mean $I_1$, $I_2$, $I_3$, and $I_4$ currents of the highlighted 1100 and 0010 input. While the $I_4$ currents are similar, the $I_1$, $I_2$, and $I_3$ currents are easily separable.*

The voltage input measurements are shown schematically in Figure 2a. Input sequences were provided in four timesteps with a 150 ms duration (Δt in Figure 1d). A -1 V pulse can be applied during the first 50 ms of each timestep. Next, the current is measured for 100 ms, always without applied voltage. The four timesteps allow sixteen different voltage input sequences. These inputs can be represented as binary numbers, where a timestep with an applied -1 V pulse is denoted as a binary "1", and a timestep without applied voltage as a binary "0".

An example measurement with a 1011 input sequence (input **u**(t) in Figure 1a and b) is shown in Figure 2b. Three -1 V pulses are applied to the device, with a missing -1 V pulse at the second timestep, shown as a dotted line. The currents collected in each timestep, corresponding to the readout states **x** in Figure 1a and b, are referred to as $I_1$, $I_2$, $I_3$, and $I_4$ in the plot. The current is increased after each pulse and decays in the absence of an applied voltage, consistent with an ionic displacement current. Each of the sixteen possible four-bit sequences was measured 100 times. The separability of the sixteen inputs was investigated by comparing the means and standard deviations of the $I_1$, $I_2$, $I_3$, and $I_4$ currents.

Figure 2c shows the obtained mean and standard deviation of the $I_4$ currents. As expected for an ionic displacement current, the $I_4$ current increases as a larger number of voltage pulses is applied, and when pulses are applied for later bits. Some input sequences lead to similar currents, such as the highlighted 1100 and 0010 sequences. Even so, Figure 2d shows that the mean $I_1$, $I_2$, and $I_3$ currents are easily separable. The mean currents after each bit of all inputs are given in Figure S3. Even though the standard deviations of several means overlap, the standard errors of the means are small, as demonstrated by Figure S4. This implies that the means are well-defined.

For in-sensor reservoir computing applications, it is important that values from each distribution are meaningfully different. This difference can be determined from the overlap of the probability mass of the distributions, i.e. their overlap coefficient.[26] The overlap coefficients represent the fraction of common random samples of two distributions. Large overlap coefficients indicate that it is likely that similar currents will be obtained for different inputs. The overlap coefficients of the $I_4$ current distributions are given in Figure S5a. While 68 of the 120 overlap coefficients are negligible, below 1%, 39 inputs have significant overlap coefficients of 10% or higher. These inputs could be confused if only one sample is provided to the network, potentially reducing its accuracy. The overlap of the distributions can be reduced significantly by mapping the inputs to both the $I_2$ and the $I_4$ currents, as demonstrated by Figure S5b. In this case, 106 overlap coefficients are below 1%, and only 9 overlap coefficients above 10% are found. Figure S5c shows that mapping to all four currents results in negligible overlap coefficients for all inputs. Nonetheless, an important caveat is that mapping to multiple currents requires additional memory elements in the in-sensor computing array, increasing device complexity and reducing energy efficiency.

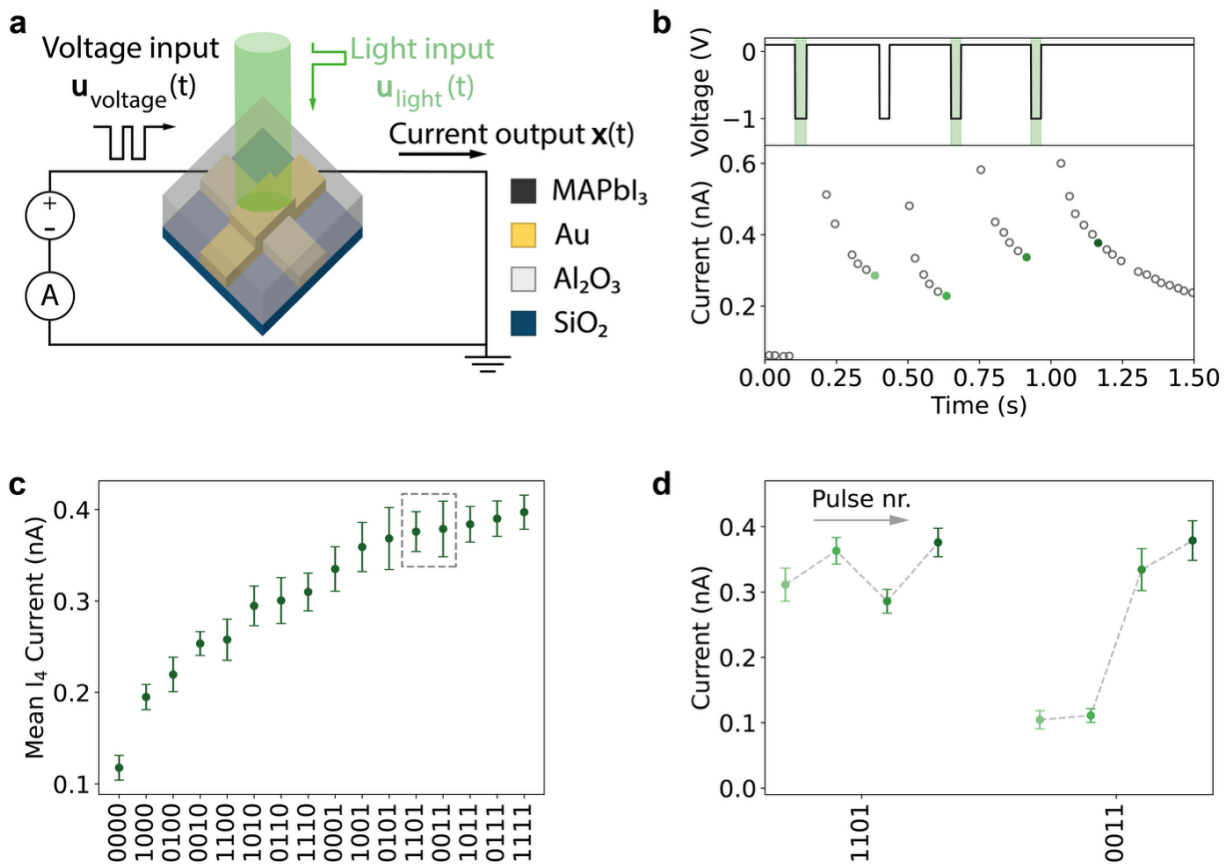

*Figure 3. Optoelectronic measurements of four-bit light pulse inputs. **(a)** Schematic of the measurements. Periodic -1 V pulses are applied to the device. Four-bit light input sequences are applied simultaneously with the -1 V pulses. **(b)** An example measurement of the 1011 sequence is shown as green shaded regions in the top panel. Measured currents*

*are shown in the bottom panel. The $I_1$, $I_2$, $I_3$, and $I_4$ currents used for further analysis are shown in green. **(c)** Mean $I_4$ currents for each four-bit input. The mean currents are higher for inputs where more light pulses are applied, and where these pulses are applied for later bits. The gray dotted box highlights the 1101 and 0011 inputs as an example of inputs with similar means. **(d)** The $I_1$, $I_2$, $I_3$, and $I_4$ currents of the 1101 and 0011 inputs are highlighted in **(c)**. While the $I_4$ currents are similar, the $I_1$, $I_2$, and $I_3$ currents are noticeably different.*

Next, we investigate the separability of the input sequences using light in addition to voltage pulses. A schematic of the light input measurement is given in Figure 3a. Each timestep (Δt in Figure 1d) for the light inputs was 270 ms. At the start of each timestep, a -1 V pulse is applied for approximately 40 ms. Afterwards, the voltage is changed to a constant +100 mV, and the current is measured. The voltage pulses set the input frequency (Δt) of the device. Light pulses can be applied simultaneously with the -1 V pulses. Applying a light pulse during the -1 V pulse is considered a binary "1", while no applied light during the -1 V pulse is represented by a binary "0". The device was not illuminated at any other part of the timestep. In our current implementation, it was not possible to induce a volatile current by applying light pulses without a bias voltage. This suggests that illumination does not generate a photovoltage that can drive ion migration. Fabricating devices with electronically asymmetric electrodes could remedy this limitation.[19]

An example measurement of a 1011 input sequence (**u**(t) in Figure 1a and b) is shown in Figure 3b. Similar to the example voltage input measurement in Figure 2b, the current is increased after each -1 V pulse. When the device is illuminated during the voltage pulse, the current increase is enhanced, in accordance with our previous work.[22] This enhancement can be explained by the higher ion mobility in the perovskite layer under illumination. Interestingly, the current also seems to decay more slowly for the light inputs compared to the voltage input in Figure 2b. Fits to the current decay after the 0000 and 0001 input sequences in Figure S6 show that this is due to a shift from a faster drift to a slower diffusion decay. This trend indicates that accumulated ions experience a weaker electric field after applying light pulses. A possible explanation might be a flattening of the electronic bands as the perovskite layer is illuminated, due to the increase in electronic charge carrier density.[27] Analogous to the voltage sequences, the separability of each of the sixteen possible inputs was investigated based on the means and standard deviations of the $I_1$, $I_2$, $I_3$, and $I_4$ currents (**x** in Figure 1b).

The means and standard deviations of the $I_4$ currents for each input sequence are given in Figure 3c. Higher currents are obtained when a larger number of light pulses are applied, and when the pulses are applied in later timesteps. Similarly to the voltage inputs, some sequences for the light inputs show comparable $I_4$ current. The constant 100 mV offset we use reduces the overlap somewhat. Figure S7 shows the mean $I_4$ currents if no offset was applied, resulting in more similar values. Nonetheless, several means are closely spaced, such as those of the highlighted 1101 and 0011 input sequences. Despite these similar mean $I_4$ currents, the $I_1$, $I_2$, and $I_3$ currents are more easily separable, as demonstrated by Figure 3d. The mean $I_1$, $I_2$, $I_3$, and $I_4$ currents and standard deviations of each 4-bit light sequence are given in Figure S8. Similar to the 4-bit voltage sequences measurements, the standard errors of the means are small, as illustrated by Figure S9. Overlap coefficients are given in Figure S10. Of the 120 coefficients, 55 are 10% or larger when mapping to only the $I_4$ current. High overlap coefficients are found particularly for inputs containing three or four light pulses. Similar to the voltage inputs, overlap is reduced significantly by mapping to both $I_2$ and $I_4$, with only 18 significant overlap coefficients. Again, mapping to all four currents yields no significant overlap for any combination of inputs.

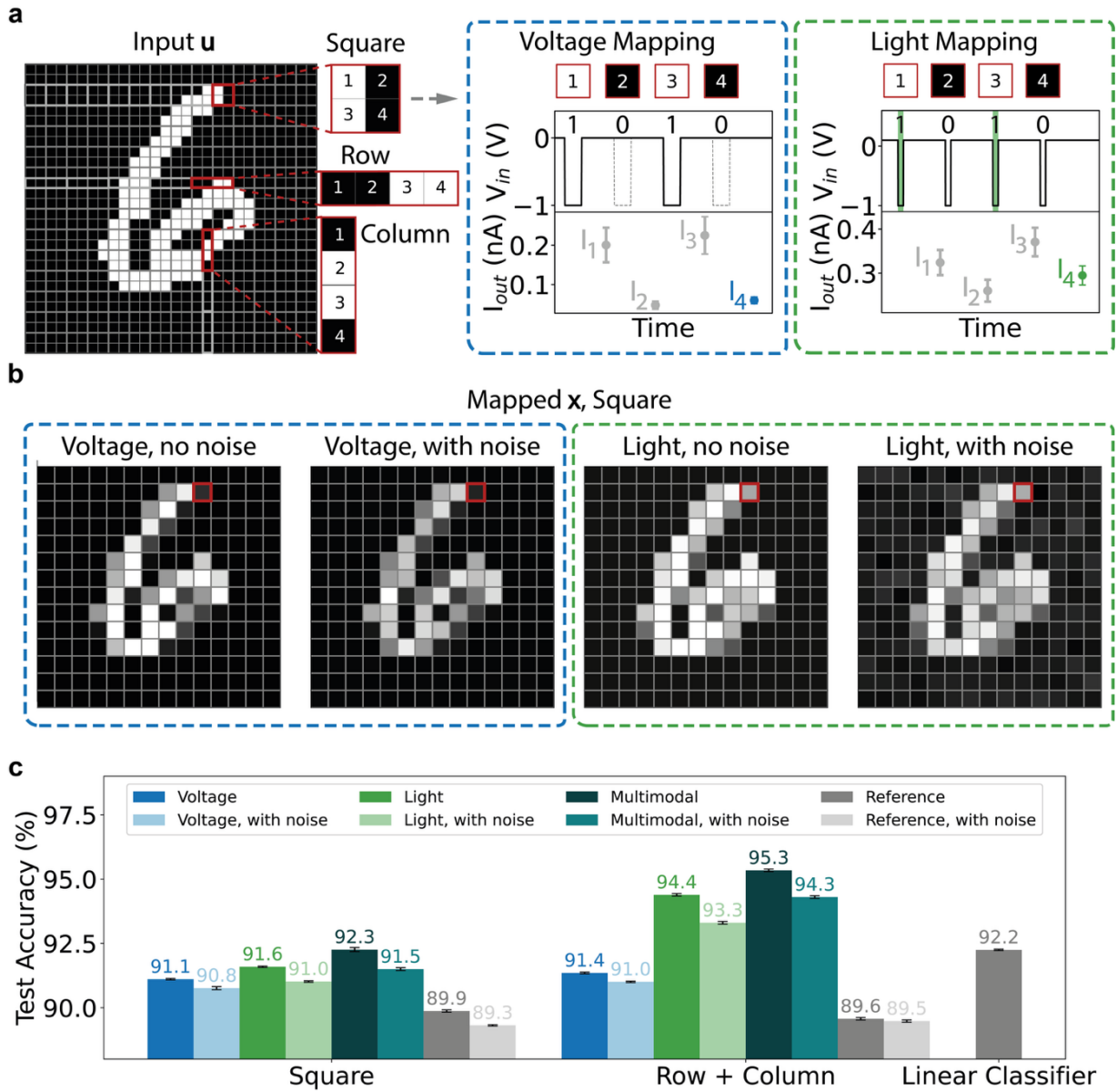

*Figure 4. MNIST classification with images transformed based on light and voltage inputs. (a) Example transformation of a binarized MNIST image of a number 6. The image is divided into 2x2 pixel squares ("Square"), 4 pixel rows ("Rows"), or 4 pixel columns ("Columns"). Pixels are laid out in a row from the pixels labeled "1" to "4". White pixels are interpreted as a "1" (voltage or light pulse input), and black pixels as a "0" (no input). The obtained 4-bit binary sequence is matched to experimental voltage or light inputs, shown for the 1010 sequence of the square example. The square, row, or column is converted (mapped) to the $I_4$ current of that sequence. (b) Example transformation based on square mapping to the voltage and light-input data. In the "no noise" transformations, the 2-by-2 pixel squares were mapped to the mean $I_4$ current of the sequences. Noise was included in the "with noise" case as described in the Methods section. The pixel highlighted in red corresponds to the red patch in (a). (c) Mean classification accuracies of datasets transformed by mapping the images with the Square, or combined Row and Column approach to the $I_4$ currents of the voltage, light, or a combination of both (multimodal) measurements. Accuracies are compared to references for which each square, row, or column was mapped to the binary value of pixel 4 in (a), without considering the values of pixels 1, 2, and 3. From this, we can determine the benefit of the volatile memory (see Methods for details). The accuracies are compared to a linear classifier reference trained on the binarized MNIST dataset without any transformations. Each test accuracy was determined from 10 independent runs with different random seeds. Error bars indicate one standard deviation.*

In the next step, we demonstrate the use of our optoelectronic artificial synapse in a reservoir network. Reservoir networks were implemented based on either the voltage or light sequences. MNIST handwritten digit classification was used to benchmark network performance. Each sample of the binarized MNIST dataset was divided into 2x2 pixel patches ("Square"), 4 pixel rows ("Row"), or 4 pixel columns ("Column"). These 4-pixel arrays were then converted to 4-bit binary sequences. White pixels are interpreted as a "1" (an input voltage or light pulse) and black pixels as a "0" (no input pulse). The sequence was constructed by combining the obtained binary values in the order denoted in Figure 4a. Next, we define a mapping $f: A \rightarrow B$, where A is any of the 16 possible 4-bit sequences, obtained from the image, and B is the corresponding $I_4$ current from Figure 2c (voltage input) or Figure 3c (light input). An example mapping of a 2x2 pixel square is shown in Figure 4a. The square is converted to the 1010 pixel sequence, which is mapped to the $I_4$ current of the 1010 voltage (Figure 2c) or light input (Figure 3c). In a physical implementation, each square, row, or column would contain the input to one device in the sensor array. The 4-bit sequences would then be applied as in Figure 2b or Figure 3b, and afterwards the $I_4$ current of each device in the array would be collected for readout. This method of temporally encoding segments of an image is commonly used in reservoir computing.[28] The purpose of the reservoir in this application is to correlate the pixel values to extract features in the images important for their classification. The *Square*, *Row*, and *Column* approach to transforming the images will therefore lead to the extraction of features in one (*Column* and *Row*) or two dimensions (*Square*).

Figure 4b shows the MNIST sample after mapping all squares to the corresponding $I_4$ currents of the 4-bit voltage and light inputs. The transformed images correspond to the mapped vector **x** in Figure 1b, where the $I_4$ currents represent the states collected from each node. We account for experimental noise in the transformation by mapping each square to a random number taken from a normal distribution with the experimentally determined $I_4$ current mean and standard deviation for that sequence. Brighter pixels, corresponding to higher $I_4$ currents, are obtained for the light-transformed images compared to voltage-inputs. Thus, compared to the voltage-input based transformations, pixel values (currents) from earlier inputs are retained to a greater extent for the light-transformed images, in line with the example 1010 mapping in Figure 4a. The longer retention can be explained by the shift to slower current decay by ionic diffusion after light inputs (Figure S6), resulting in a longer memory window.

Each image in the MNIST dataset was transformed by the *Square*, *Row*, or *Column* mapping approach. Linear readout layers were then trained on the transformed datasets. Figure 4c shows the obtained classification accuracies over ten independent runs for different transformations. The networks are compared to reference networks trained on the binarized MNIST dataset, for which each square, row, or column was mapped to the binary value of pixel 4 in Figure 4a. These reference networks are equivalent to using a regular, memory-less sensor array (**s**), such as a camera, in combination with a resistor array weight matrix (**W**). Comparing the reservoir network accuracies with the reference, therefore, allows an accurate assessment of the contribution of the reservoir. The accuracies are also compared to a linear classifier trained directly on the binarized MNIST dataset. Reservoir transformations successfully increase the linear separability of the dataset if reservoir network accuracies exceed that of this reference.[5]

For the *Square* mapping approach, the light-based networks outperformed those based on voltage inputs (0.5%, p < 0.001). We show in Supplementary Note 1 with simulations that this is due to the longer memory window of light inputs, which results in higher contrast for patches at the edges of the digits after the transformation. The ratio of the retention time to the input frequency of the light inputs is close to the optimum in the simulations. Conversely, the shorter relative memory window of the voltage-based networks provides higher contrast at patches containing many white pixels, which are typically found in the centers of the digits. This is also visualized in Figure 4b. Previously, better performance of reservoir networks was found when combining two sensing modes into a single, multimodal network.[15,29] These multimodal networks combine two nonlinear transformations to increase linear separability. We construct multimodal networks by combining the $I_4$ currents of the voltage and light networks into a single mapped state vector **x**. The weight matrix is trained on the combined states.

The multimodal network outperforms all other networks implementing the *Square* mapping, with a mean accuracy of 92.3 ± 0.1% (all differences $p < 0.001$). The improved performance likely stems from the complementary combination of enhanced contrast at the edge of the digit by the light-mapping and at the centers of the digits by the voltage-mapping. The addition of noise comparable to the experimental noise only slightly decreases the accuracy of the networks. Adding noise decreased the light-input and multimodal network accuracy more strongly, likely due to the higher overlap coefficients of the light input $I_4$ currents (Figure S10) than those of the voltage inputs (Figure S5). In all cases, the reservoir networks outperform the *Square* reference, also when considering the noise. This result shows that the transformations encode information relevant for classification, also when considering experimental noise. Nevertheless, even without considering experimental noise, the multimodal network with the *Square* transformation does not significantly outperform the linear classifier ($p = 0.681$).

In previous implementations of in-sensor reservoir networks, the MNIST dataset is typically transformed in a four-pixel line-by-line fashion ("Row", or "Column" in Figure 4a).[15,30–32] We implement the same transformations in Figure S12a, b, and c. The Figure shows that these transformations distort the image more strongly than the *Square* mapping method we follow in Figure 4. The classification accuracies are given in Figure S12d. The accuracies were slightly higher for the *Row* mapping, both for the voltage (89.0 ± 0.1 %, 88.6 ± 0.1% with noise) and light inputs (90.8 ± 0.1%, 90.0 ± 0.1% with noise), but are lower than those obtained for the *Square* mapping in Figure 4c. This is likely due to a stronger loss of relevant features as the images are compressed in only one dimension. Interestingly, the mean accuracy of the multimodal network was 92.6 ± 0.1% (91.5 ± 0.1% with noise), slightly higher than for the *Square* mapping ($p < 0.001$). This accuracy exceeds that of the linear classifier ($p < 0.001$), although only if experimental noise is not considered. Similar to the *Square* mapping, this could be explained by the complementary combination of the light and voltage $I_4$ current mapping. Furthermore, the four-pixel rows extend over a larger distance in the image. This might allow the network to better extract relevant features for classification. Notably, the accuracies of the light input and the multimodal networks are higher than those reported in previous work.[15,30–32] For the light-based network, this could be due to a more favorable current output for the input sequences, as explained in Supplementary Note 1. Another explanation might be a more thorough hyperparameter search before training the readout layers. For the multimodal network, the accuracy is likely increased further by the complementary combination of the light and voltage mapping, in line with previous work.[15,29] Importantly, some previously reported accuracies are comparable or even lower than the 1 Reading reference we report here.[15] This finding highlights the importance of thoroughly evaluating the reservoir performance in comparison to reference linear classifiers to prevent overestimating the contribution of the reservoir transformations.

A proven way to increase the accuracy of a reservoir computing network is to present the images to the network at different rotations.[28] We implement this here by mapping each digit with both the *Row* and *Column* approaches and concatenating the obtained mapped image vectors to obtain a single combined vector **x** for each image. The readout layers are then trained on the combined *Row* and *Column* transformed dataset. Network accuracies are shown in Figure 4c as "Row + Column". As expected, mean accuracies increased compared to those of the *Square* and separate *Row* and *Column* mapping. The highest obtained mean accuracy was 95.3 ± 0.1% for the multimodal network. By combining the *Row* and *Column* mappings, the separability of features is increased in both dimensions. Compared to the *Square* mapping, features are extracted over larger distances in the images (4 instead of 2 pixels in either direction), which can explain the better performance. Both the light-input and multimodal networks implementing this mapping approach exceed the linear classifier. These networks also further improve on previously reported in-sensor reservoir implementations.[15,30–32] Increasing the number of pixels mapped by each device could further enhance the network accuracy.[28] The light inputs in particular appear to have a sufficiently long retention time for mapping more than the currently implemented 4 pixels at a time. Previously, applying additional rotations to the MNIST dataset resulted in higher accuracies for software reservoir networks.[28] The same approach might increase the accuracy of our networks as well. Another method to increase the accuracy is to add hidden layers that perform further nonlinear transformations of the data.[19] However, such are difficult to implement in hardware and are therefore less interesting for in-sensor computing.

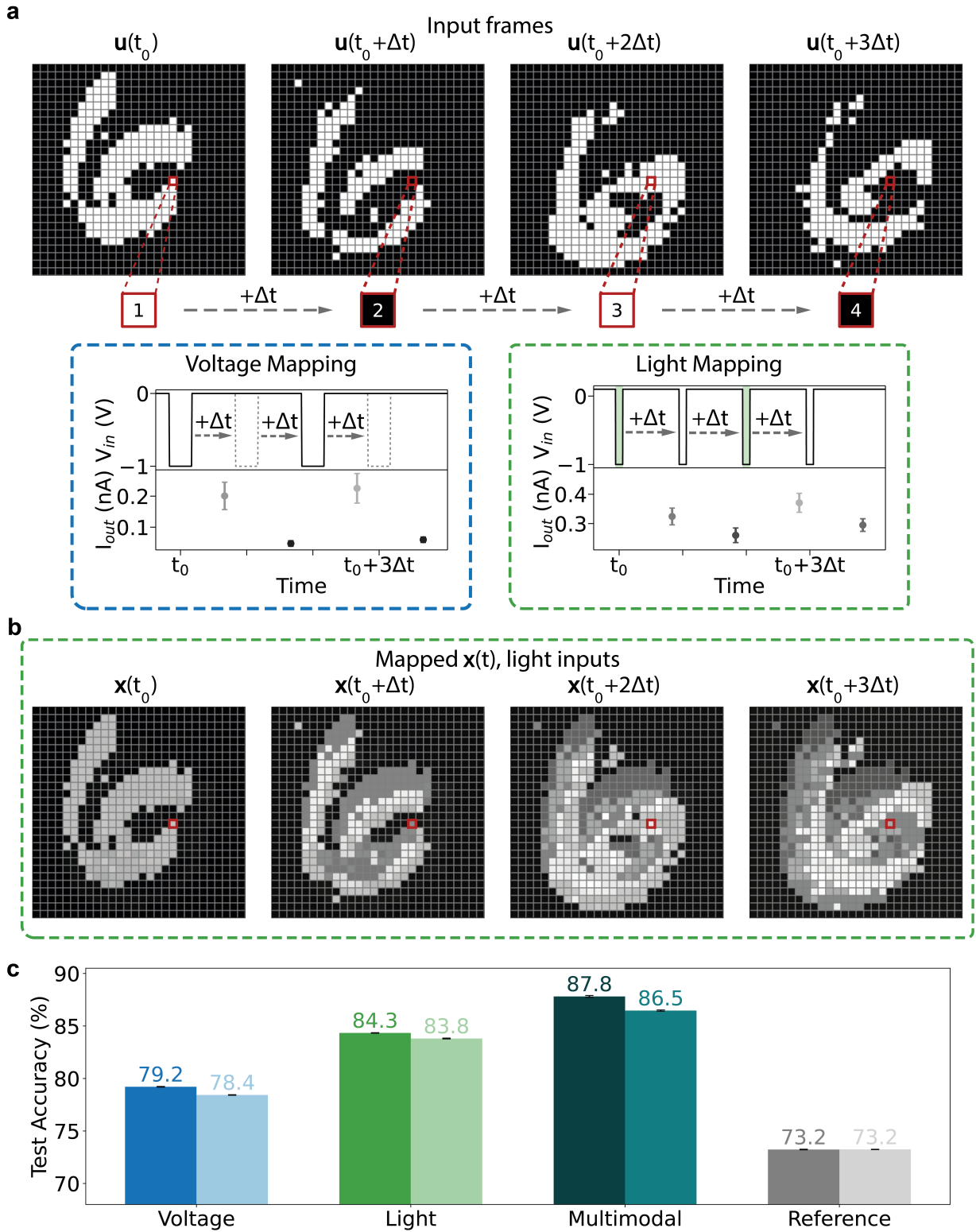

*Figure 5. Handwritten digit classification from video with frames transformed based on light and voltage inputs. **(a)** Four consecutive frames of a number 6. Pixels underneath each frame show the pixel highlighted in red at each timestep. For each pixel in the four frames, its four consecutive values are converted to a four-bit light or voltage pulse input sequence. The sequence is then applied (in silico) to the experimentally measured artificial synapse. **(b)** Transformations of the frames in **(a)** after each timestep, based on the means of the four-bit light inputs. Each pixel is mapped following the procedure outlined in **(a)**. The red boxes highlight the same pixel as in the frames in **(a)**. **(c)** Classification accuracies obtained for the reservoir networks based on voltage (blue), light (green), and multimodal (teal) inputs. The accuracies are compared to a reference (gray) trained on the dataset without transformations. Accuracies for networks also considering experimental noise are given in lighter colors. Accuracies were obtained by taking the mean of 10 independent training runs.*

Next, we investigated the reservoir network performance for natively temporal inputs relevant to in-sensor computing. Handwritten digit classification from video, based on the N-MNIST dataset,[33] is taken as an example. While the light input lends itself well to detecting and transforming visual data, the voltage implementation could be useful for processing data from a separate sensor detecting, for example, tactile signals.[15] We followed a similar approach as before. First, 34 by 34 pixel binned and binarized N-MNIST frames are mapped to the experimental voltage and light data, as shown schematically in Figure 5a. However, instead of squares, rows, or columns, individual pixels of four consecutive frames are mapped to the $I_1$, $I_2$, $I_3$, and $I_4$ currents.

Figure 5b shows the frames after each pixel is mapped to the corresponding mean currents of the 4-bit light sequence. The number six of the previous frames remains visible with lower brightness in each consecutive frame, visualizing that information from previous frames is retained. Figure S13 shows the transformed sample accounting for the experimental noise, and the sample transformed based on voltage inputs. In all cases, the number six remains recognizable after the transformation. However, the features of previous frames remain brighter for the sample mapped to the light input. This is again a result of the longer retention of information compared to the voltage input, similar to Figure 4.

The dataset was transformed based on voltage and light input experimental currents. As in Figure 4, multimodal networks were implemented by combining the voltage and light-input transformed frames. Next, linear readout layers were trained on the transformed datasets. A reference network was trained on a dataset that was not transformed. In our implementation, the frames (**u**(t)) are projected on the sensor array (**s**), and the output currents of each device (**x**(t)) are collected continuously. Weighted sums of the currents are taken with the weight matrix (**W**) after each input frame (for each timestep Δt), to classify each frame consecutively. Analogous to the MNIST classification in Figure 4, the reference network is equivalent to an in-sensor network with a conventional, memory-less sensor array such as a camera. The benefit of the transformations can be evaluated by comparison with the reference. Network accuracies are given in Figure 5c. The Supplementary Movies show example transformations and predictions by the networks. Compared to the MNIST dataset, the video data is significantly less linearly separable, as follows from the lower reference classification accuracy of 73.2 ± 0.1%. Interestingly, markedly higher classification accuracies were found for the reservoir networks (79.2 ± 0.1%, 84.3 ± 0.1%, and 87.8 ± 0.1% for voltage, light, and multimodal sequences, respectively, all $p < 0.001$). This implies that the transformations substantially increased the linear separability of the dataset. The better performance of the light compared to the voltage input networks we find here is again likely due to the longer retention time of the light inputs (see Supplementary Note 1). The higher accuracy of the multimodal network again likely follows from the complementary transformations by the light and voltage networks. Also, similar to the MNIST classification results, adding experimental noise resulted in only a slight decrease in accuracy (0.8%, 0.5%, and 1.3% for the voltage, light, and multimodal networks, respectively, all $p < 0.001$).

The considerable increases in classification accuracy of the MNIST handwritten digits from both images and video show that the volatile optoelectronic devices are promising for in-sensor computing applications. Particularly when implementing them in multimodal networks. Moreover, the back-contacted microscale device architecture lends itself well to high-density integration in in-sensor computing arrays. Currently, we have only implemented binary inputs. The capabilities of these arrays could be further extended by considering different light intensities and voltage magnitudes. This extension would allow processing real-time, analog signals. Analog inputs could simultaneously increase the range of accessible reservoir states, potentially resulting in even more capable nonlinear transformations.

# Conclusions

In summary, we have demonstrated physical reservoir networks based on back-contacted, microscale halide perovskite devices. The networks encoded MNIST images and N-MNIST-based videos based on experimentally measured currents of the device using 4-bit voltage and light input sequences. We have shown with drift-diffusion simulations that the measured current transients are due to ion migration in the device. When employed in a network in silico, multimodal networks based on the devices that processed both light and voltage inputs gave notably higher classification accuracies compared to a linear classifier reference. The mean accuracy of the multimodal network was 95.3 ± 0.1%. We explain with simulations that the high accuracy is due to light-based transformations complementing those based on voltage due to their difference in retention time, emphasizing different features in the images. As a result, the accuracies we report here for the light-based and multimodal networks are higher than those in previous work that followed a similar approach.[15,30–32] Notable accuracy increases with respect to the reference were found for MNIST classification from video as well. The mean accuracy of the networks increased by up to 14.6% to 87.8 ± 0.1% for the multimodal mapping. Hence, the ability of the halide perovskite devices to process both voltage and light inputs allows multimodal processing in the same chip, significantly improving accuracy. These accuracy gains, combined with the microscale device architecture that lends itself well to high-density integration, are promising for efficient in-sensor computing applications.

# Materials and Methods

**Materials**
Silicon wafers with a 100 nm dry thermal oxide layer were purchased from Siegert Wafer. $PbI_2$ (99.99%) was purchased from TCI. Methylammonium iodide (MAI, purity not listed) was purchased from Solaronix. $Al(CH_3)_3$ (97%), anhydrous DMF, DMSO, and chlorobenzene were purchased from Sigma-Aldrich. MA-N1410 resist and MA-D533/s developer were purchased from Micro Resist. All materials were used without further purification.

**Halide perovskite device fabrication**
Halide perovskite devices were fabricated as reported previously.[22] In short, 2.5 μm wide gold bottom electrodes with a thickness of 80 nm were patterned on the silicon wafer using a lift-off process with the MA-N1410 resist. A 15 nm $Al_2O_3$ insulating layer was deposited over the bottom electrode by atomic-layer deposition from $Al(CH_3)_3$ and $H_2O$ precursor gases in a home-built atomic-layer deposition setup at 250 °C. The lift-off process was repeated to pattern 2.5 μm wide, 80 nm thick gold top contacts perpendicular to the bottom electrodes.

Inside a N$_2$-filled glovebox, the MAPbI$_3$ precursor was prepared by dissolving 1.1 mmol PbI$_2$ and MAI in 1 mL DMF and 0.1 mL DMSO. The precursor was filtered through a 0.2 μm PTFE filter and spin-coated on a die cut from the patterned wafer at 4000 rpm for 30 seconds. After 5 seconds, 250 μL of chlorobenzene was added to the spinning sample. The sample was annealed at 100 °C for 10 minutes. The sample was then encapsulated with Blufixx epoxy and a glass coverslip, cured for 1 minute with a UV torch.

**4-bit input measurements**

The 4-bit input sequence measurements were performed with a Keysight B2902A Precision Source/Measure Unit. One channel of the SMU was used to apply voltage pulses to the device and measure the output current. For the light inputs, a second channel was used to drive a 520 nm high-power Cree XLamp XP-E LED. The irradiance was 2.8 mW/cm$^2$, measured with a Thorlabs PM100D optical power meter with an S120VC sensor. Measurements of each input sequence were repeated 100 times. The $I_1$, $I_2$, $I_3$, and $I_4$ currents in the main text were determined by taking the mean and standard deviation over all measurements.

**Drift-diffusion simulations**

The drift-diffusion simulations were carried out using Setfos by Fluxim with the device parameters listed in Table 1. We simulated the current due to mobile ions after applying a voltage pulse train of 1 to 4 voltage pulses of -1 V.

*Table 1: Simulation parameters used for the drift-diffusion simulations. A schematic of the simulated device stack is given in Figure S2c and d.*

| Parameter | Value | Comment |
|---|---|---|
| Al$_2$O$_3$ Thickness (nm) | 15 | |
| Relative permittivity insulator | 9 | |
| Electron affinity insulator (eV) | 2.5 | |
| Band gap insulator (eV) | 5 | |
| Thickness perovskite (nm) | 50 | |
| Relative permittivity perovskite | 24.1 | Taken from ref 34 |
| Electron affinity perovskite (eV) | 3.9 | Taken from ref 35 |
| Band gap perovskite (eV) | 1.6 | Taken from ref 35 |
| Effective density of states perovskite conduction band (1/cm$^3$) | 8x10$^{18}$ | |
| Effective density of states perovskite valence band (1/cm$^3$) | 8x10$^{18}$ | |
| Mobile positive ion density (1/cm$^3$) | 3x10$^{17}$ | |
| Immobile negative ion density (1/cm$^3$) | 3x10$^{17}$ | |
| Mobility mobile ions (cm$^2$/Vs) | 1x10$^{-10}$ | |
| Work function electrodes (eV) | 5.1 | |
| Voltage pulse duration (s) | 0.05 | |
| Time between voltage pulses (s) | 0.1 | |

**Transformations of the MNIST and N-MNIST datasets**

Lookup tables were constructed from the experimentally measured mean currents of each 4-bit input. Means were taken over 100 measurements for each input. The 28-by-28-pixel MNIST images were binarized with a threshold of 0.5. The binarized images were then divided into square 2-by-2-pixel patches, 4-pixel rows, or 4-pixel columns. The pixels of each patch were converted to a 4-bit sequence, as shown in Figure 4a. The sequences were then mapped by matching them to a 4-bit input sequence in the lookup table. The corresponding currents were added to a new array representing the transformed image. To account for the experimental noise, a random number was taken from a normal distribution with the mean and standard deviation determined from the 100 measurements of the corresponding $I_4$ current, instead of mapping to the $I_4$ current mean. The mean and standard deviation of each sequence are displayed in Figure 2c (for voltage inputs) and Figure 3c (for light inputs). For the references, a Gaussian blur was applied with a variance determined from the original MNIST dataset, see Figure S14a. This is the noise that would be expected if the binarized MNIST images are projected on a conventional camera. Figure S14b shows that the blurring (kernel size = 5, σ = 0.522) approximates the observed noise well.

The N-MNIST dataset was imported with Tonic (version 1.6.0). The original spiking, event-based dataset was filtered, binned, and thresholded to reconstruct videos of the original moving MNIST images. This modified dataset is more relevant for real-time video recognition using the in-sensor networks for simultaneous detection and processing. A denoise filter with a 10 ms filter time was applied to the dataset, and the samples were binned into 50 ms frames. Next, the pixel values in the first four 34-by-34-pixel frames of each sample were normalized, and the frames were binarized with a threshold of 0.2. This threshold yielded the most recognizable digits in the frames. A similar approach was followed to map the frames to the voltage and light-input data as for the MNIST mapping. However, instead of squares, rows, or columns, each individual pixel was mapped based on its four consecutive values in the four frames. After mapping each pixel, the four frames were added to the dataset separately. This extended the training and test datasets from 60,000 to 240,000 and from 10,000 to 40,000 samples, respectively. Noise was introduced as Gaussian blurring with the same parameters as for the MNIST dataset.

**Network training**

Readout layers were trained on the input vectors with PyTorch (version 2.6.0), using the Adam optimizer. Hyperparameters (learning rate, weight decay, and $β_1$ and $β_2$ of the Adam optimizer) were tuned over 100 runs with Optuna (version 4.2.1), using the Tree-structured Parzen Estimator algorithm. A large, fixed batch size of 256 was chosen for faster training.[36] The MNIST dataset was randomly split into a *training* set containing 50,000 samples and a *validation* set of 10,000 samples during hyperparameter tuning. Readout layers with 1960 (14x14x10, *Square,* or 7x28x10, *Row*, and *Column* mapping), 3920 (7x28x10x2, combined *Row* and *Column* mapping, or 14x14x10x2, multimodal *Square* mapping), or 7840 (7x28x10x2x2, multimodal combined *Row* and *Column* mapping) weights were trained on the mapped images. The N-MNIST dataset was randomly split into a 200,000-sample *train* and a 40,000-sample *validation* set. The readout layers, consisting of 11560 (34x34x10), or 23120 (34x34x10x2, multimodal mapping) weights, were trained on the transformed frames.

After hyperparameter tuning, training was repeated on the full training dataset with the optimal hyperparameters. Mean network accuracies and standard deviations were recorded from a seed sweep over the same 10 seeds for all networks. All mean accuracies and their standard deviations are rounded to the first decimal place. Standard deviations that would be rounded to 0.0% (e.g. 0.03%) were rounded up to 0.1% instead to account for experimental error. All standard deviations were ≤ 0.10 %.

Paired t-tests were performed to check the significance of differences in accuracy. All mean accuracies within each plot were significantly different ($p < 0.001$), with the exception of the accuracies of the multimodal, *Square* mapping network and the reference in Figure 4c ($p = 0.681$), and the reference networks with and without noise for the *Row* mapping in Figure S12d ($p = 0.275$), and the video dataset in Figure 5c ($p = 0.546$).

# Acknowledgements

The work of J.J.B., A.O.A., M.C.S., and B.E. received funding from the European Research Council (ERC) under the European Union's Horizon 2020 research and innovation programme under grant agreement No. 947221. The work is part of the Dutch Research Council NWO and was performed at the research institute AMOLF. The authors thank Marc Duursma, Bob Drent, Igor Hoogsteder, Arthur Karsten, and Laura Juškėnaitė for technical support.

# Declaration of interests

The authors declare no competing interests.

Supplementary Information to

# Microscale multimodal reservoir networks of halide perovskite

Jeroen J. de Boer[1], Agustin O. Alvarez[1], Moritz C. Schmidt[1], Bruno Ehrler[1,*]

[1]*LMPV-Sustainable Energy Materials Department, AMOLF, 1098 XG, Amsterdam, the Netherlands*
*\*Correspondence: b.ehrler@amolf.nl*

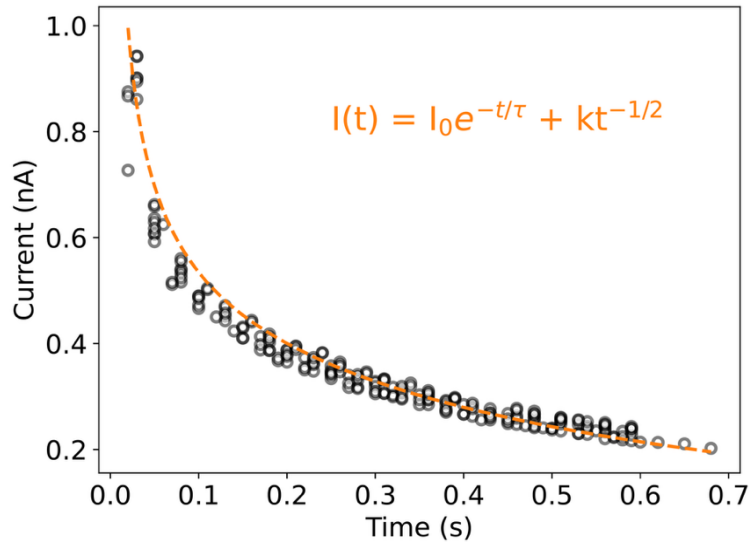

Figure S1. *Fit to the combined data of ten transient current measurements as in Figure 1d. The current is measured after the final -1 V pulse, and fit with equation* $I(t) = I_0 e^{-t/\tau} + kt^{-1/2}$. *Fitting parameters are given in Table S1.*

Table S1. *Fitting parameters for the fit in Figure S1, along with the fitting error.*

| Fitting parameter | Value |
|---|---|
| $I_0$ | 0.22 ± 0.02 nA |
| $\tau$ | 0.53 ± 0.04 s |
| k | 0.110 ± 0.005 nA $\sqrt{s}$ |

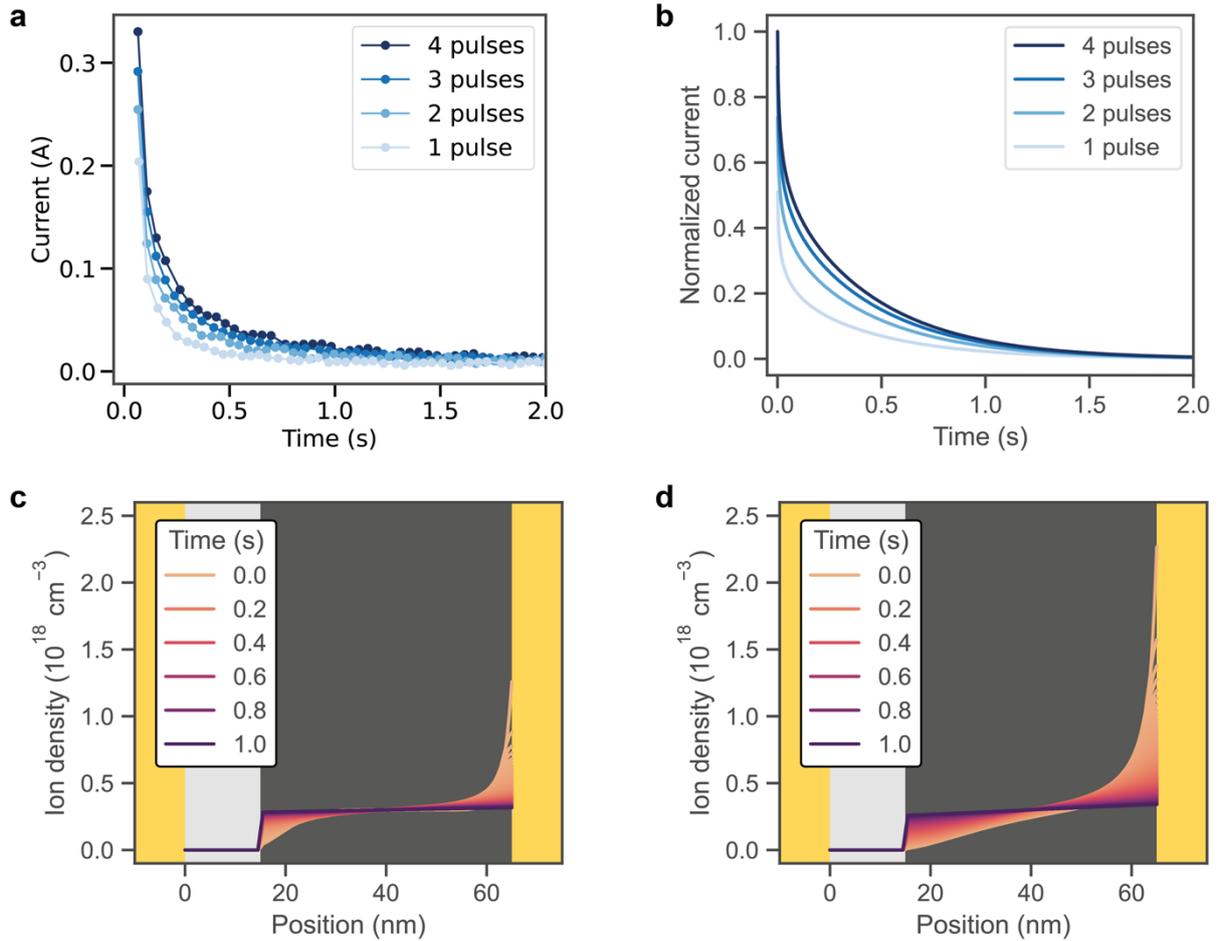

Figure S2. Comparison of measured transient currents with drift-diffusion simulations. **(a)** Measured currents after applying a series of -1 V pulses. The final pulse is removed at t = 0 sec. **(b)** Drift-diffusion simulations of the measurements. **(c)** Ion densities after applying a single -1 V pulse. Ions are initially accumulated at the cathode at t = 0 sec and redistribute throughout the perovskite layer within seconds. **(d)** The same simulation as in **(c)**, but after applying four -1 V pulses. The same effect is observed, but with a larger magnitude.

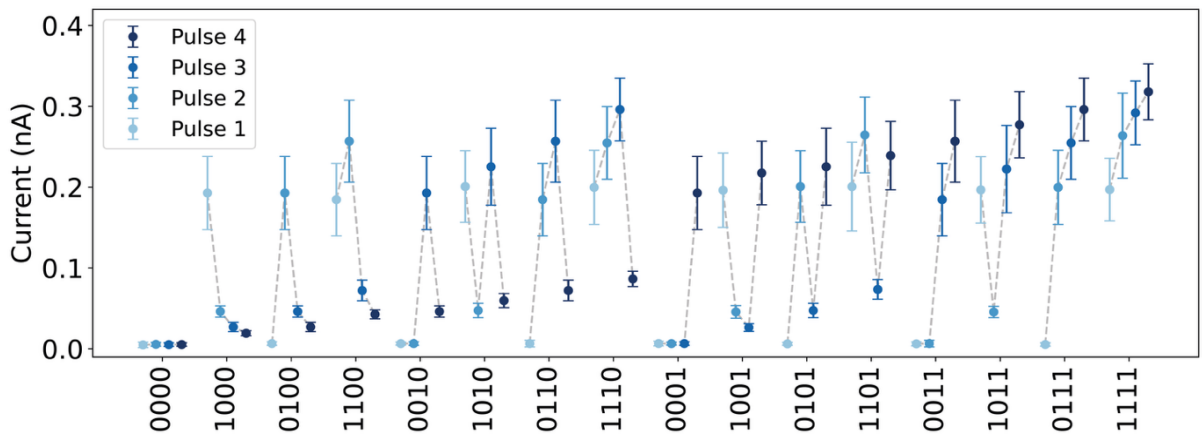

Figure S3. Mean currents after each of the four pulses of the 4-bit voltage inputs. Means taken over 100 measurements, with error bars representing one standard deviation. Gray dotted lines are added for each input to guide the eye. Inputs with similar mean final currents are more easily separable by considering intermediate currents.

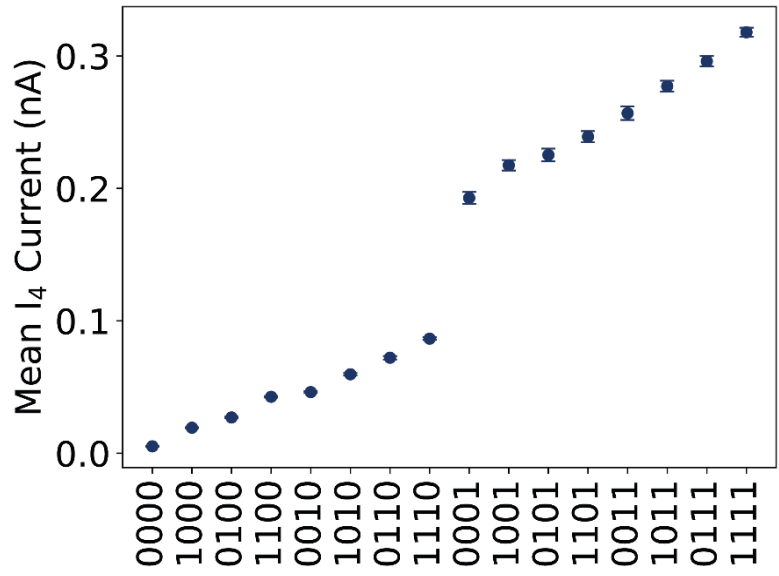

*Figure S4. Mean currents with error on the mean of the 4-bit voltage inputs. The low errors on the means indicate that the obtained means are well-defined.*

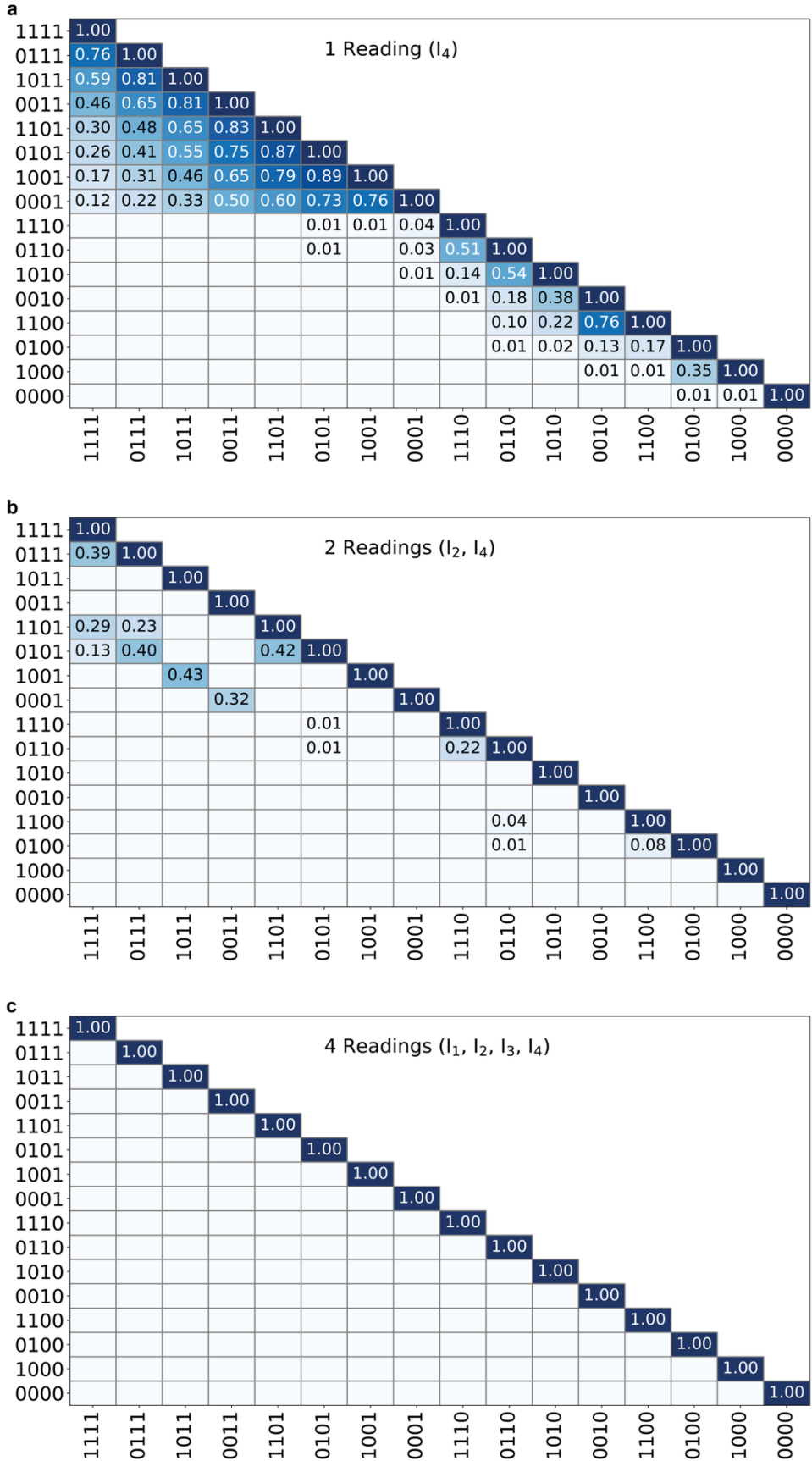

*Figure S5. Overlap matrices of the four-bit voltage input mappings to **(a)** the current after the fourth bit, **(b)** the current after the second and the fourth bit, and **(c)** the currents after all four bits. Overlap coefficients below 0.01 are not plotted for clarity. Significant overlap coefficients are found for similar inputs when mapping to fewer currents.*

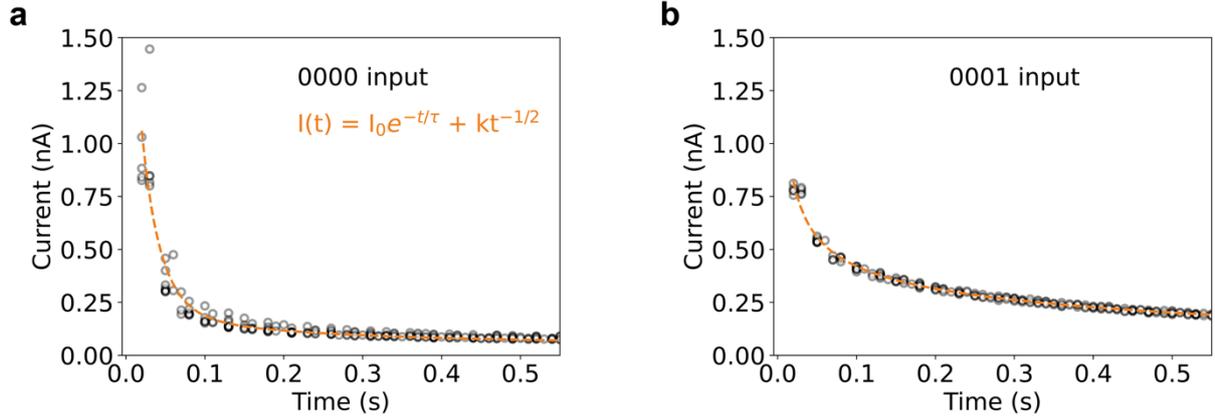

*Figure S6. Fits to the combined data of ten 0000 **(a)** and 0001 **(b)** light input sequence transient current measurements. The current is measured after the final -1 V pulse and fit with equation $I(t) = I_0 e^{-t/\tau} + kt^{-1/2}$. Fitting parameters are given in Table S2.*

*Table S2. Fitting parameters of the current decays in Figure S6a and b. Errors represent the fitting error.*

| Input | $I_0$ (nA) | $\tau$ (s) | $k$ (nA $\sqrt{s}$) |
|---|---|---|---|
| **0000** | 1.64 ± 0.14 | 0.023 ± 0.002 | 0.052 ± 0.003 |
| **0001** | 0.12 ± 0.01 | 0.737 ± 0.066 | 0.100 ± 0.002 |

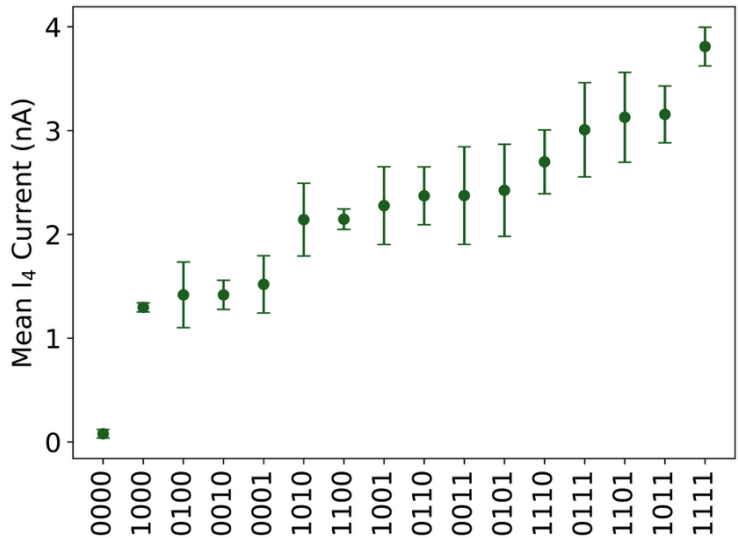

*Figure S7. Mean $I_4$ current for each of the 4-bit inputs if no 100 mV offset bias is applied between the -1 V pulses. More overlap between $I_4$ currents is found compared to the measurement with the offset in Figure 3c of the main text.*

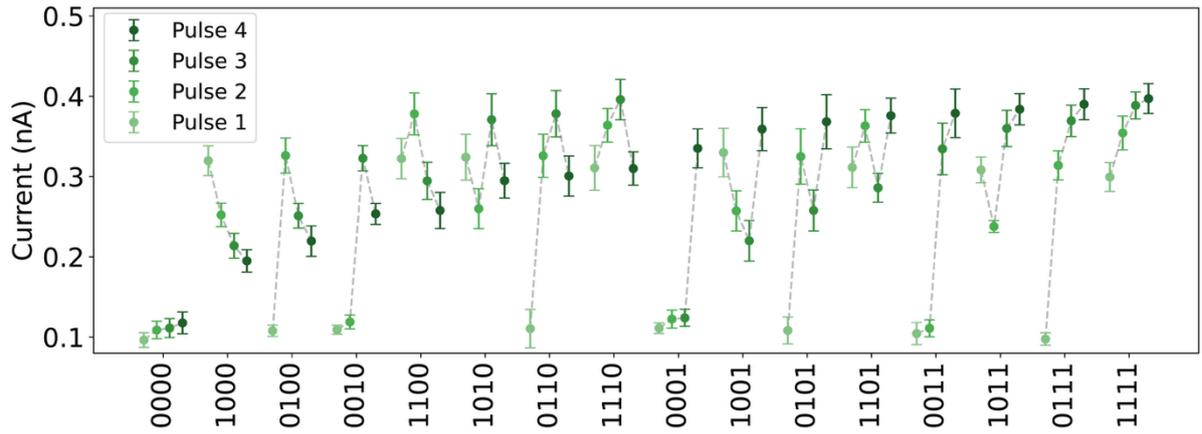

*Figure S8. Mean currents after each of the four pulses of the 4-bit light inputs. Means taken over 100 measurements, with error bars representing one standard deviation. Gray dotted lines are added for each input to guide the eye. Inputs with similar mean final currents are more easily separable by considering intermediate currents.*

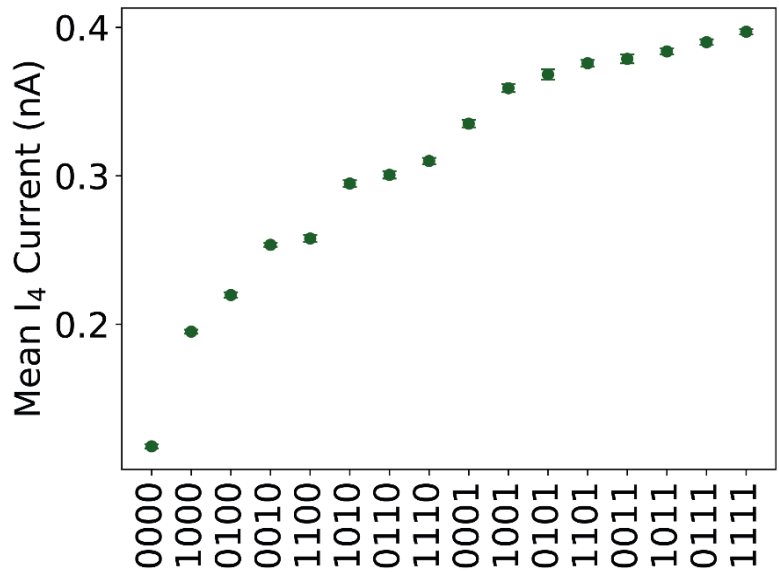

*Figure S9. Mean currents with error of the mean of the 4-bit light inputs. The low errors of the means indicate that the obtained means are well-defined.*

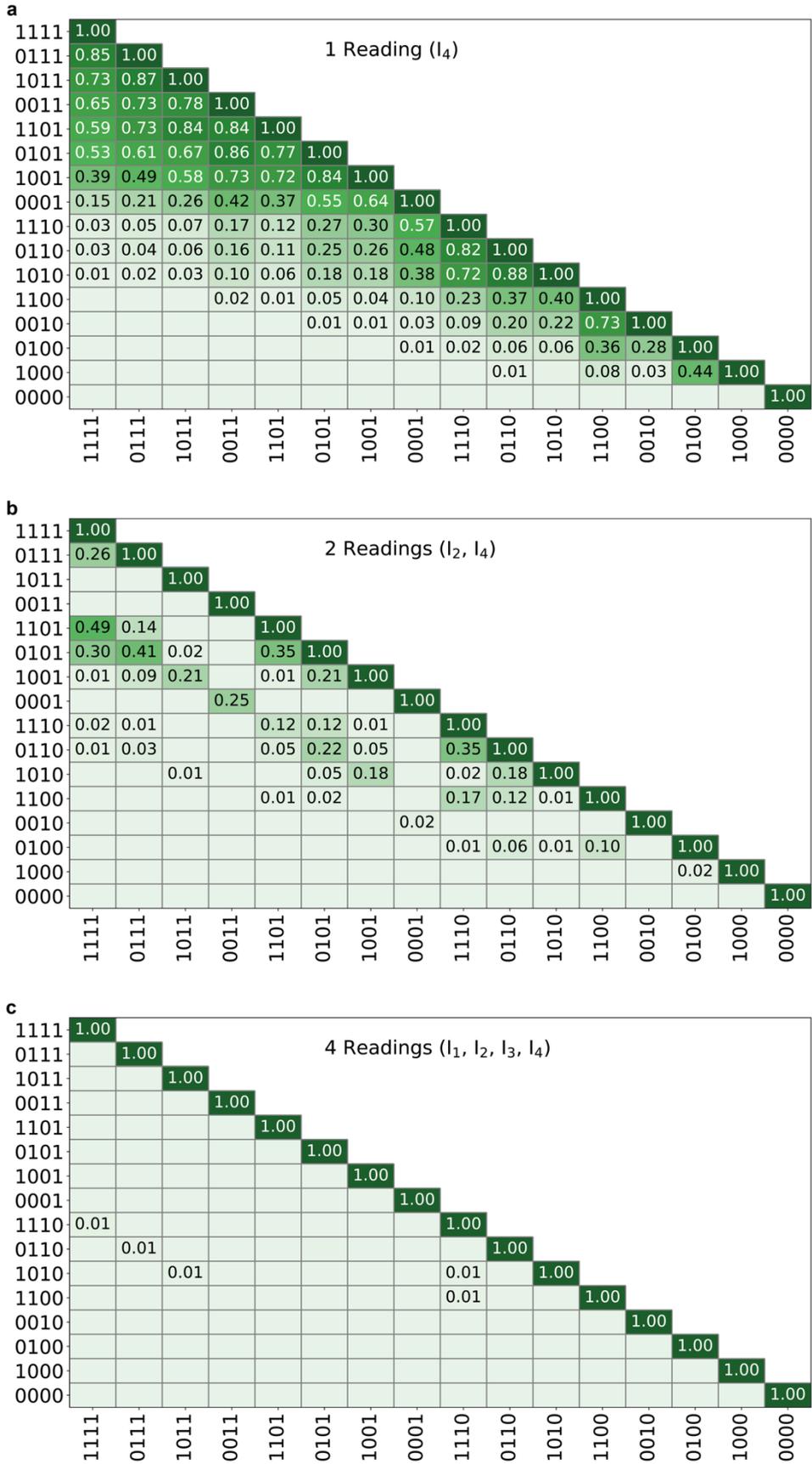

*Figure S10. Overlap matrices of the four-bit light input mappings to **(a)** the current after the fourth bit, **(b)** the current after the second and the fourth bit, and **(c)** the currents after all four bits. Overlap coefficients below 0.01 are not plotted for clarity. Significant overlap coefficients are found for similar inputs when mapping to fewer currents.*

# Supplementary Note 1. Impact of the I₄ currents on classification accuracy

This Supplementary Note investigates the impact of the obtained currents for 4-bit sequences on the MNIST classification accuracy. We simulate I₄ currents and show the effect of different trends on the MNIST image transformations to explain differences in classification accuracy. The I₄ currents obtained for the light (Figure 3c) and voltage inputs (Figure 2c) are compared to the generated I₄ currents to explain the differences in performance of the networks in the main text. Finally, we discuss how plotting the I₄ currents for different 4-bit input sequences allows facile estimation of the in-sensor reservoir network performance. These results can be used to find optimal experimental parameters.

In reservoir computing, the reservoir performs a nonlinear transformation of the input,[1] increasing the separability of otherwise convoluted features.[2] In our implementation, the reservoir consists of devices that transform 4-bit voltage and light input sequences into a current. In this context, input sequences are separable if distinct current values are obtained for each sequence. This requires the device to have a volatile memory. Because of its memory, the device outputs a higher I₄ current for a 0011 sequence than it does for a 0001 input. The volatility, on the other hand, causes the output current to depend on the order in which light or voltage pulses are applied in the sequence. It extends the device response beyond an additive, linear response in which inputs with the same number of pulses, such as the 1001 and 0011 sequences, yield the same I₄ current. Due to the current decay in our experiments, lower I₄ currents are output if a pulse is applied earlier in the sequence. Consequently, a lower I₄ current is obtained for the 1001 sequence than for the 0011 input. The volatility is thus crucial for separating inputs.

Ideally, the I₄ currents should be as distinct as possible after the transformation. The overlap between I₄ currents is minimized if they follow a linear trend in plots such as those in Figure 2c and 3c in the main text. Simulated I₄ currents following a linear trend are given as the gray dotted line in Figure S11a. The normalized current is calculated as $I_4(s) = \frac{s}{N-1}$, where I₄ is the current of the s$^{th}$ sequence on the x-axis and $N$ is the number of possible sequences (16 sequences for a 4-bit input). The first sequence (0000) is assigned $s = 0$. The 4-bit voltage and light sequences in the main text produced I₄ currents that deviate from this linear trend. The voltage inputs (Figure 2c) yielded lower I₄ currents for the first half of the sequences. The light inputs (Figure 3c), on the other hand, produced I₄ currents above the linear trend over the whole range of sequences. We simulate these responses with polynomial functions.

I₄ currents falling below the linear trend are calculated as $I_4(s) = \left(\frac{s}{N-1}\right)^n$, shown in blue in Figure S11a. In the plot, n is 1.5, 2.0, 3.0, or 4.0, where higher values result in a stronger convex, parabolic shape. These trends compress I₄ currents of earlier sequences on the x-axis (the 0000 end) and stretch out later ones (on the 1111 end). In experimental measurements of the I₄ current, this is due to high volatility, causing a 0000 and 1000 input, for example, to give a similar output. Currents above the linear trend were calculated as $I_4(s) = 1 - \left(1 - \frac{s}{N-1}\right)^n$ (red curves in Figure S11a), using the same range of n-values. These currents follow a concave parabolic curve, with a stronger curvature for higher values of n. In this case, currents for later sequences on the x-axis are compressed, and those of earlier inputs are stretched. In experimental measurements, this response is caused by long retention times, resulting in a saturation of the $I_4$ current (e.g. a similar $I_4$ current for 0111 and 1111 sequences).

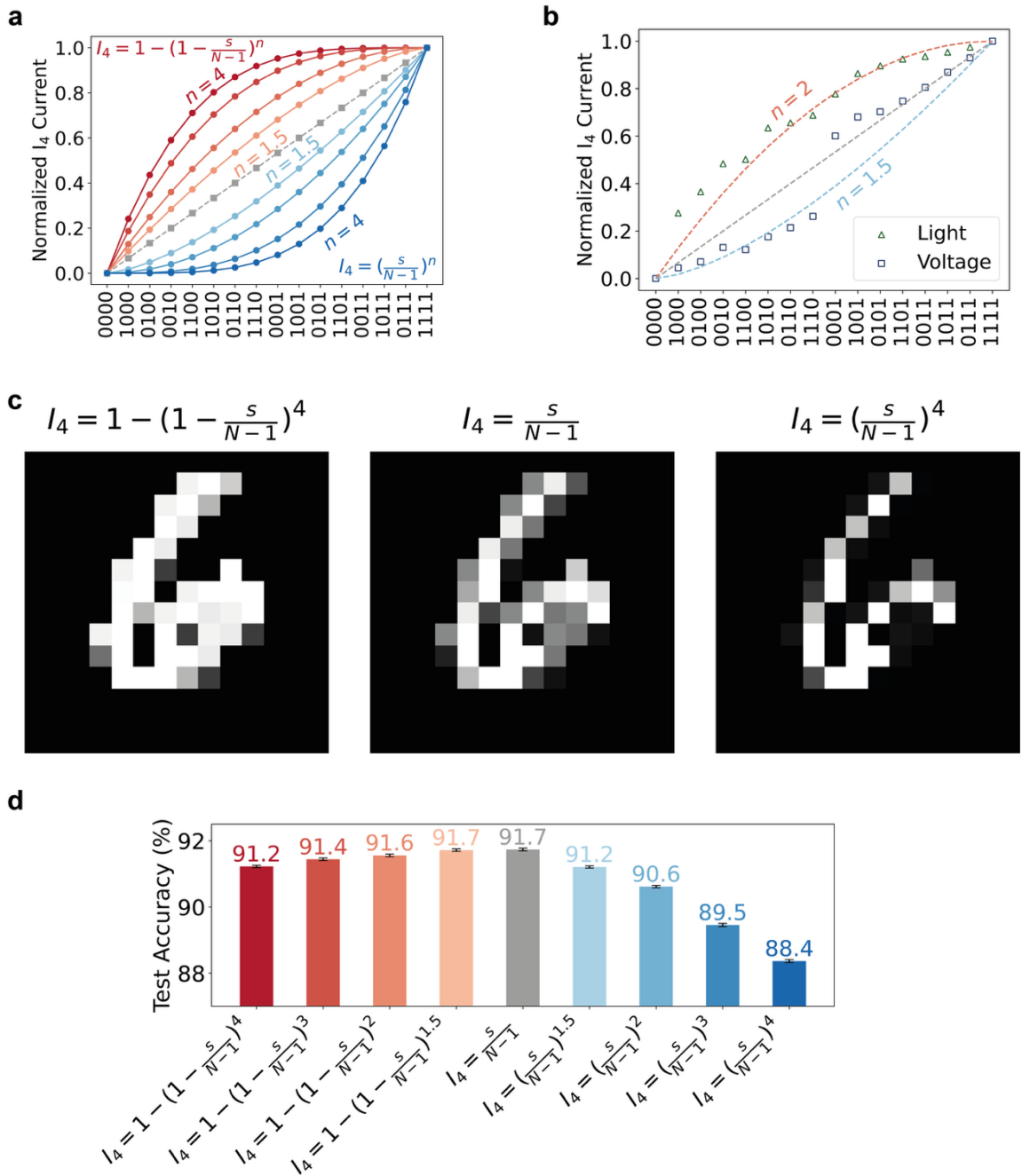

Figure S11. Transformations of the MNIST dataset based on simulated $I_4$ current following linear and polynomial trends. **(a)** Input sequences with normalized $I_4$ currents that increase linearly (squares and dotted gray line), or according to polynomial functions $I_4(s) = 1 - \left(1 - \frac{s}{N-1}\right)^n$ (red circles and continuous lines) and $I_4(s) = \left(\frac{s}{N-1}\right)^n$ (blue circles and continuous lines), with $n \in \{1.5, 2, 3, 4\}$. **(b)** Normalized $I_4$ currents of the light (green triangles) and voltage (blue squares) inputs compared to the artificial $I_4$ currents. The light input $I_4$ currents follow a similar trend to the polynomial $I_4(s) = 1 - \left(1 - \frac{s}{N-1}\right)^2$. The $I_4$ currents of the voltage inputs up to the 1110 sequence resemble those of the function $I_4(s) = \left(\frac{s}{N-1}\right)^{1.5}$ and become more linear afterwards. **(c)** The same MNIST image as in Figure 4a mapped to the $I_4$ currents calculated based on the $I_4(s) = 1 - \left(1 - \frac{s}{N-1}\right)^4$, $I_4(s) = \frac{s}{N-1}$, and $I_4(s) = \left(\frac{s}{N-1}\right)^4$ functions. Mapping to the $I_4(s) = \frac{s}{N-1}$ currents results in a number six with a range of grayscale values. Mapping to currents generated by the polynomial

functions yields more extreme grayscale values, with more bright pixels for $I_4(s) = 1 - \left(1 - \frac{s}{N-1}\right)^4$ and more dark pixels for $I_4(s) = \left(\frac{s}{N-1}\right)^4$. **(d)** Classification accuracies of the datasets mapped to the linear and polynomial functions. The highest accuracy of 91.7% is obtained for the linear relation. Minor accuracy penalties are obtained for the concave $I_4(s) = 1 - \left(1 - \frac{s}{N-1}\right)^n$ functions, while they are more severe for the convex $I_4(s) = \left(\frac{s}{N-1}\right)^n$ functions.

These polynomial functions were chosen because they stretch and compress the linear response to the same extent, but at opposite ends of the x-axis. The equal magnitude of stretching and compression allows a fair comparison of the two trends.

Figure S11b illustrates that the measured $I_4$ currents of the first half of the voltage and all light-inputs are described best by $I_4(s) = \left(\frac{s}{N-1}\right)^{1.5}$ and $I_4(s) = 1 - \left(1 - \frac{s}{N-1}\right)^2$, respectively.

The effects of the deviations from the linearly increasing $I_4$ currents are visualized in Figure S11c. The linearly increasing $I_4$ currents result in a transformation of the MNIST image to an image with a range of grayscale values. Transformations based on convex ($I_4(s) = \left(\frac{s}{N-1}\right)^4$) or concave ($I_4(s) = 1 - \left(1 - \frac{s}{N-1}\right)^4$) $I_4$ currents result in more extreme grayscale values. Convex $I_4$ currents force intermediate values in the image transformed based on linear $I_4$ currents closer to 0, resulting in a lossy transformation of the image. The lower right part of the loop of the number six, for example, is no longer visible in the image. Conversely, the image transformed based on concave $I_4$ currents force otherwise intermediate values closer to 1. This results in saturation in the transformed image, for example in the upper right part of the loop of the number six. At the same time, features that were more difficult to notice, around the curved line of the number for example, are more distinguishable for this transformed image. Both the lossy and saturated transformations are expected to decrease later classification accuracies, as different features become more difficult to distinguish.

Figure S11d shows the classification accuracies obtained by linear readout layers trained on the transformed datasets. The linearly increasing $I_4$ currents give the highest accuracy (91.7 ± 0.1%). The mean accuracies for concave $I_4$ current transformations decreased only slightly. For n = 1.5, no statistically significant difference was found compared to the linear $I_4$ transformation (p = 0.252). For higher values of n, the differences were statistically significant (p < 0.001), but modest, with an accuracy of 91.2 ± 0.1% for n = 4. Interestingly, the accuracy penalty was markedly more severe for the convex $I_4$ currents (p < 0.001 for all differences). The same mean accuracy of 91.2 ± 0.1% for the n = 4 concave $I_4$ currents was reached for n = 1.5 already. The accuracy then decreased further for larger values of n, down to 88.4 ± 0.1% for n = 4. The accuracies for the convex n = 1.5 (91.2 ± 0.1%) and concave n = 2 (91.6 ± 0.1%) $I_4$ transformations are well-matched to those obtained for the voltage (91.1 ± 0.1%) and light (91.6 ± 0.1%) inputs in Figure 4c. This indicates that the polynomial functions are good approximations for the experimentally measured $I_4$ currents.

These results suggest that, for MNIST classification, the transformation should ideally give equal separability for all inputs (linear trend), or slightly emphasize the earlier sequences in Figure S11 where fewer pulses are applied (mildly concave trends). These inputs correspond to patches with few white pixels, which typically constitute the edges of the digits. Our results, therefore, suggest that these edges should remain distinguishable from the background for optimal classification. Saturation of inputs with more applied pulses, due to transformations with concave $I_4$ currents, impacts the classification accuracy less severely. Contrast between patches with many white pixels, therefore, seems less important for classification.

Our findings explain the better performance of the light-input networks in Figure 4c and 5c. Moreover, the simulations are generalizable for MNIST classification with any 4-bit input in-sensor reservoir network. This is particularly valuable when optimizing experimental parameters. $I_4$ currents, or equivalent outputs, can be measured for the different sequences and plotted as in Figure S11a. Figure S11a and d can then function as a simple lookup table to estimate classification accuracy. This removes the need for time-consuming hyperparameter tuning and training of readout layers. Timesteps of the input can be increased (for a more convex $I_4$ trend) or decreased (for a more concave $I_4$ trend) to optimize the accuracy of the network. Our results show that, at least for MNIST classification based on 4-bit inputs, experimental parameters should be adjusted to obtain linear or slightly concave $I_4$ current trends for best results. We note that the timesteps could not be decreased further for our voltage inputs to obtain more concave trends due to limitations of our experimental setup.

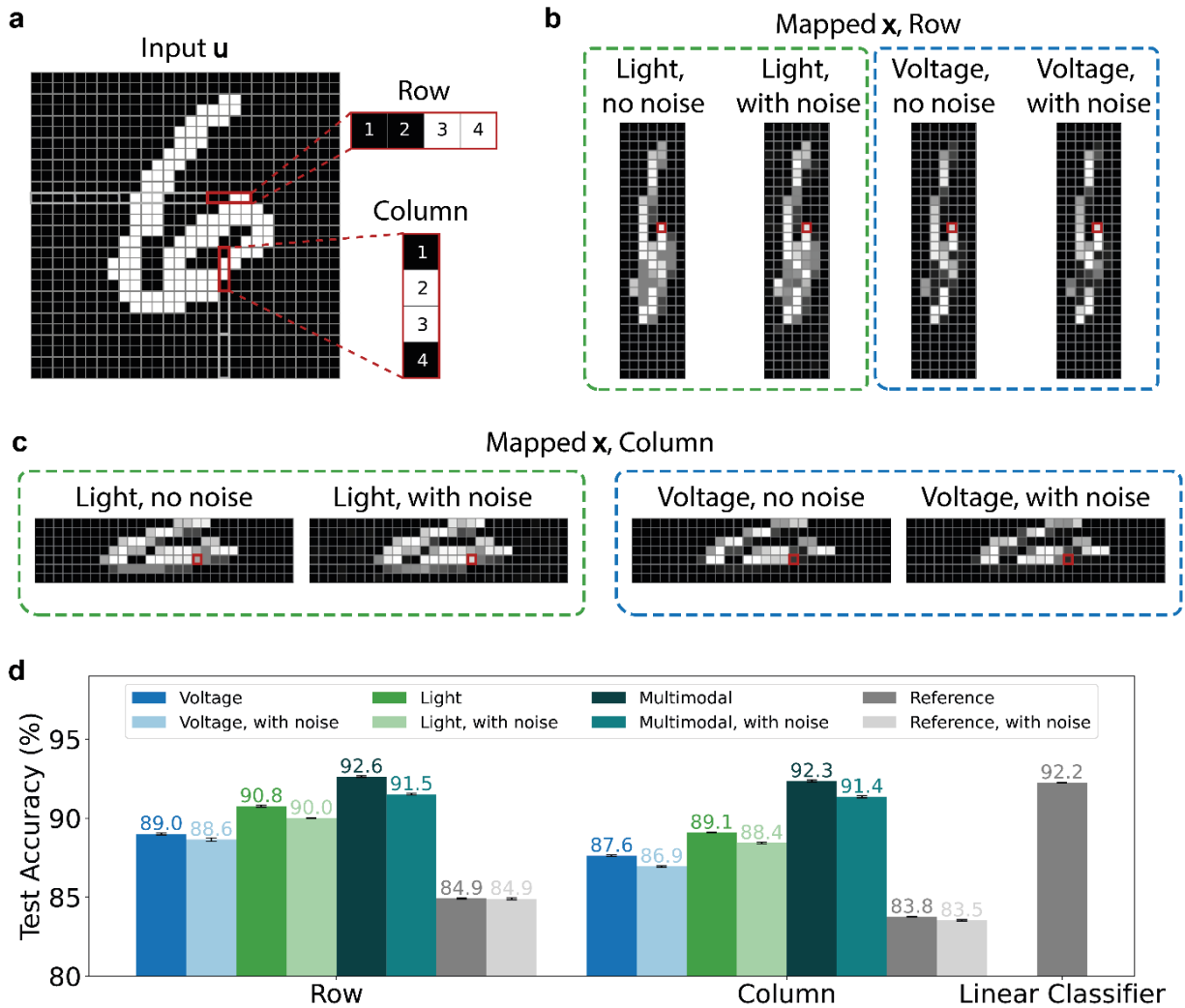

*Figure S12.* Row *and* Column *mapping for MNIST transformation and classification.* **(a)** *The same sample number six as in Figure 4a divided into four-pixel rows or columns. Rows or columns are mapped to the $I_4$ currents corresponding to their 4-bit sequences.* **(b)** *The number six transformed by* Row*-mapping. The image retains the same number of pixels vertically (28), but becomes fourfold smaller horizontally (to 7 pixels wide) by the transformation. The pixel outlined in red corresponds to the row outlined in* **(a)**. **(c)** *The number six transformed by* Column*-mapping. The image retains the same number of pixels horizontally (28), but becomes fourfold smaller vertically (to 7 pixels high) by the transformation. The pixel outlined in red corresponds to the column outlined in* **(a)**. **(d)** *Mean classification accuracies of the voltage, light, and multimodal networks for datasets transformed by* Row *and* Column *mapping. The accuracies are compared to references, for which each row or column was mapped to the binary value of pixel 4. For the voltage, light, and multimodal networks, noise was included by taking random samples from a normal distribution of the $I_4$ currents for each sequence. For the reference network, a Gaussian blur with a kernel based on the original MNIST dataset (see Methods for details) was applied to obtain a noisy dataset. The accuracies are compared to a linear classifier trained on the binarized MNIST dataset without any transformations. Each test accuracy was determined from 10 independent runs with different random seeds. Error bars indicate one standard deviation.*

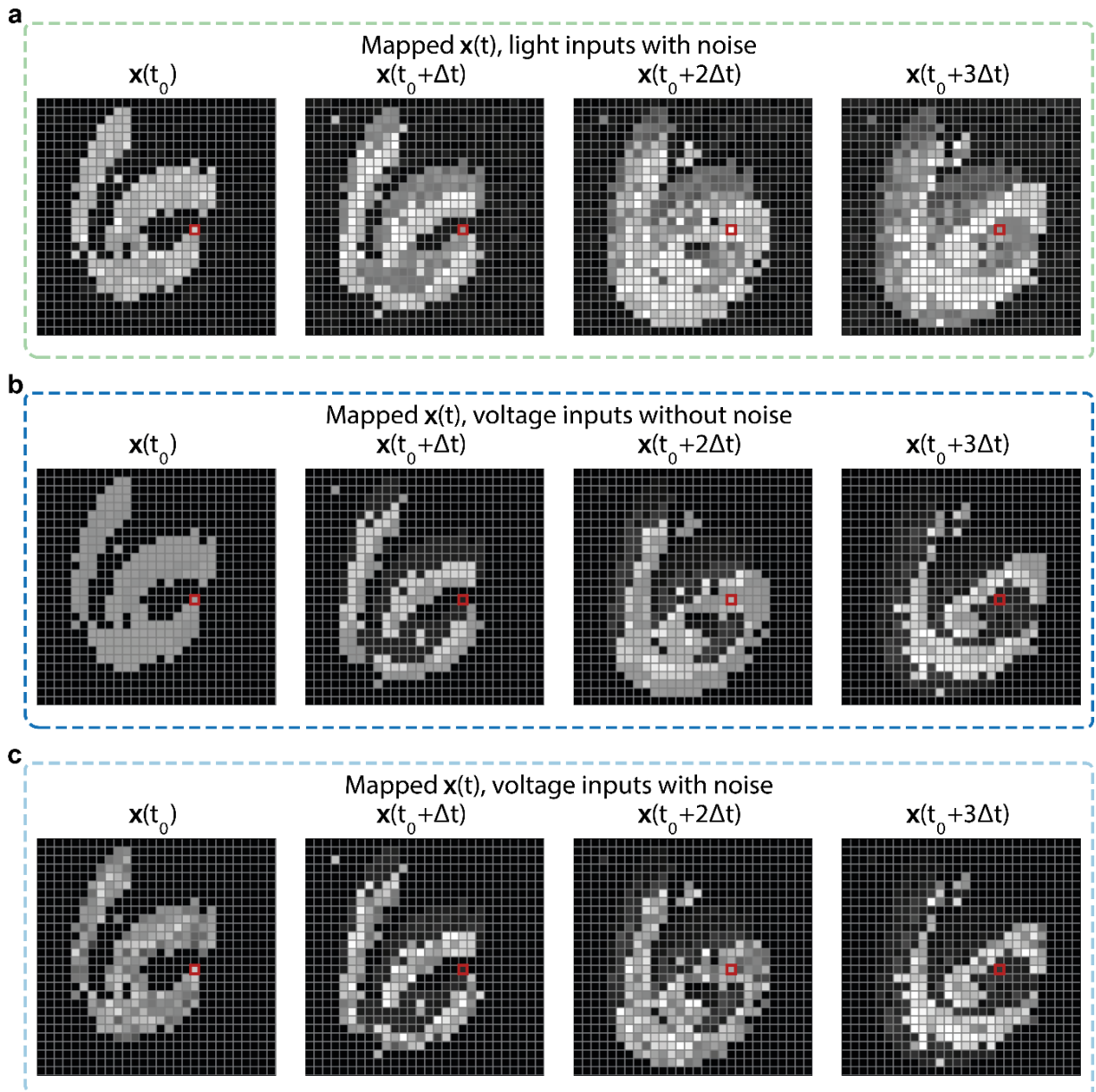

*Figure S13. **(a)** Mapping of each pixel of the N-MNIST frames to values drawn from a normal distribution with a mean and standard deviation of the corresponding 4-bit light input data. **(b)** Mapping to the mean currents obtained for the voltage input measurements. **(c)** Mapping to random values drawn from a normal distribution with means and standard deviations obtained for the corresponding 4-bit voltage inputs.*

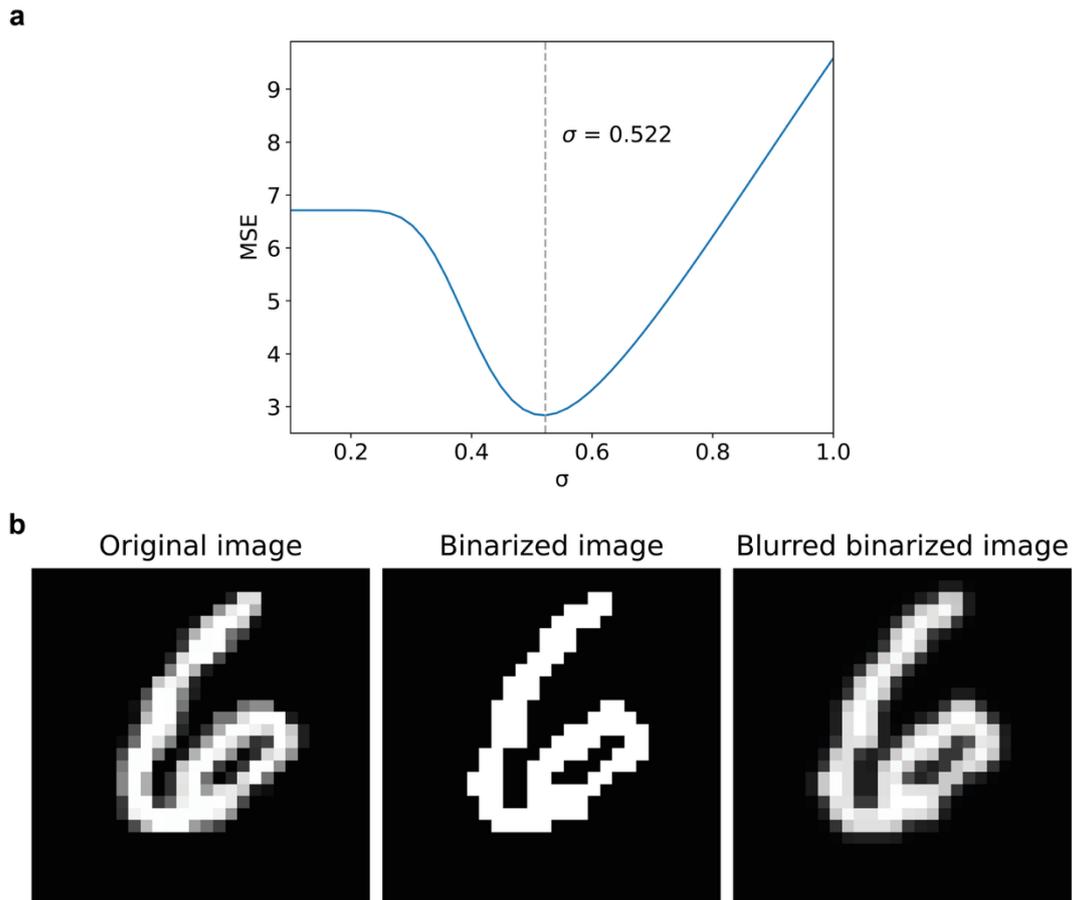

*Figure S14. Fitting of a Gaussian blurring kernel to the binarized MNIST data. **(a)** Mean squared error (MSE) obtained by comparing Gaussian blurred binarized MNIST images to the original MNIST dataset as a function of the variance (σ). The variance of the Gaussian blurring kernel with a kernel size of 5 pixels was varied between 0 and 1 to find the lowest MSE value (σ = 0.522). **(b)** Comparison of an original MNIST image (left) to the binarized image (middle) and the image obtained by applying the Gaussian blurring filter with the optimal variance to the binarized image (right).*